\documentclass[twocolumn,letter]{article}

\usepackage{url}
\usepackage{color}
\usepackage{fullpage}

\newcommand{\ignore}[1]{}

\begin{document}
\title{FactSheets: Increasing Trust in AI Services\\ through Supplier's Declarations of Conformity}

\author{
M. Arnold,$^1$
R. K.\ E.\ Bellamy,$^1$
M. Hind,$^1$ S. Houde,$^1$
S. Mehta,$^2$ A. Mojsilovi\'c,$^1$
\\
R. Nair,$^1$ 
K. Natesan
Ramamurthy,$^1$ 
D. Reimer,$^1$ 
A. Olteanu,\thanks{A. Olteanu's work was done while at IBM
Research. Author is currently affiliated with Microsoft Research.}  
\hspace{0.5em}D. Piorkowski,$^1$
\\
J. Tsay,$^1$
and K. R.\ Varshney$^1$
\\
IBM Research \\
$^1$Yorktown Heights, New York, $^2$Bengaluru, Karnataka
}

\date{}

\maketitle

\begin{abstract}
Accuracy is an important concern for suppliers of artificial
intelligence (AI) services, but considerations beyond accuracy, such
as safety (which includes fairness and explainability), security, and
provenance, are also critical elements to engender consumers' trust in
a service. Many industries use transparent, standardized, but often
not legally required documents called supplier's declarations of
conformity (SDoCs) to describe the lineage of a product along with the
safety and performance testing it has undergone. SDoCs may be
considered multi-dimensional fact sheets that capture and quantify
various aspects of the product and its development to make it worthy
of consumers' trust.  Inspired by this practice, we propose FactSheets
to help increase trust in AI services. We envision such documents to
contain purpose, performance, safety, security, and provenance
information to be completed by AI service providers for examination by
consumers. We suggest a comprehensive set of declaration items
tailored to AI and provide examples for two fictitious AI services in the appendix of the paper.
\end{abstract}

\section{Introduction} 
\label{sec-intro}

Artificial intelligence (AI) services, such as those containing predictive models trained through machine learning, are increasingly key pieces of products and decision-making workflows. A service is a function or application accessed by a customer via a cloud infrastructure, typically by means of an application programming interface (API). For example, an AI service could take an audio waveform as input and return a transcript of what was spoken as output, with all complexity hidden from the user, all computation done in the cloud, and all models used to produce the output pre-trained by the supplier of the service.  A second more complex example would provide an audio waveform translated into a different language as output. The second example illustrates that a service can be made up of many different models (speech recognition, language translation, possibly sentiment or tone analysis, and speech synthesis) and is thus a distinct concept from a single pre-trained machine learning model or library. 

In many different application domains today, AI services are achieving impressive accuracy. In certain areas, high accuracy alone may be sufficient, but deployments of AI in high-stakes decisions, such as credit applications, judicial decisions, and medical recommendations, require greater trust in AI services. Although there is no scholarly consensus on the specific traits that imbue trustworthiness in people or algorithms \cite{LevineBCS2018,Lee2018}, fairness, explainability, general safety, security, and transparency are some of the issues that have raised public concern about trusting AI and threatened the further adoption of AI beyond low-stakes uses \cite{VarshneyA2017,PapernotMJFCS2016}.  Despite active research and development to address these issues, there is no mechanism yet for the creator of an AI service to communicate how they are addressed in a deployed version. This is a major impediment to broad AI adoption.

Toward transparency for developing trust, we propose a {\em FactSheet}
for AI Services.
A FactSheet will contain sections on all relevant attributes of an AI service, such as intended use, performance, safety, and security. Performance will include appropriate accuracy or risk measures along with timing information. Safety, discussed in \cite{Moller2012,VarshneyA2017} as the minimization of both risk and epistemic uncertainty, will include explainability, algorithmic fairness, and robustness to dataset shift. Security will include robustness to adversarial attacks. Moreover, the FactSheet will list how the service was created, trained, and deployed along with what scenarios it was tested on, how it may respond to untested scenarios, guidelines that specify what tasks it should and should not be used for, and any ethical concerns of its use.  Hence, FactSheets help prevent overgeneralization and unintended use of AI services by solidly grounding them with metrics and usage scenarios.

A FactSheet is modeled after a
\emph{supplier's declaration of conformity} (SDoC).
An SDoC is a document to ``show that a product,
process or service conforms to a standard or technical regulation, in
which a supplier provides written assurance [and evidence] of
conformity to the specified requirements,'' and is used in many
different industries and sectors including telecommunications and
transportation \cite{NIST}.  
\ignore{The document typically specifies the creator, a unique
identification of the product or service including version or serial
number, lists of standardized tests conducted, technical information
about testing conditions and results, a declaration statement
indicating conformance to the tests, and the signature of the
responsible agent in the supplier organization.}
Importantly, SDoCs are often voluntary and tests reported in SDoCs are conducted by the supplier itself rather than by third parties \cite{ANSI}. This distinguishes self-declarations from certifications that are mandatory and must have tests conducted by third parties.  We propose that FactSheets for AI services be voluntary initially; we provide further discussion on their possible evolution in later sections.

Our proposal of AI service FactSheets is inspired by, and builds upon, recent work 
that focuses on increased transparency for datasets~\cite{gebru-2018,data-statements,HollandHNJC2018} and models~\cite{model-cards}, but is distinguished from these in that we focus on the final AI service.  We take this focus for three reasons:

\begin{enumerate}
\item AI services constitute the building blocks for many AI applications. Developers will query the service API and consume its output. An AI service can be an amalgam of many models trained on many datasets.  Thus, the models and datasets are (direct and indirect) components of an AI service, but they are not the interface to the developer.

\item Often, there is an expertise gap between the producer and consumer of an AI service. The production team relies heavily on the training and creation of one or more AI models and hence will mostly contain data scientists. The consumers of the service tend to be developers.  When such an expertise gap exists, it becomes more crucial to communicate the attributes of the artifact in a standardized way,
as with Energy Star or food nutrition labels.

\item  Systems composed of safe components may be unsafe and, conversely, it may be possible to build safe systems out of unsafe components, so it is prudent to also consider transparency and accountability of services in addition to datasets and models.  In doing so, we take a functional perspective on the overall service, and can test for performance, safety, and security aspects that are not relevant for a dataset in isolation, such as generalization accuracy, explainability, and adversarial robustness. 

\end{enumerate}
Loukides et al.\ propose a checklist that has some of the elements we seek \cite{LoukidesMP2018}. 

Our aim is not to give the final word on the contents of AI service FactSheets, but to begin the conversation on the types of information and tests that may be included.  Moreover, determining a single comprehensive set of FactSheet items is likely infeasible as the context and industry domain will often determine what items are needed. One would expect higher stakes applications will require more comprehensive FactSheets. Our main goal is to help identify a common set of properties. A multi-stakeholder approach, including numerous AI service suppliers and consumers, standards bodies, and civil society and professional organizations is essential to converge onto standards. It will only be then that we as a community will be able to start producing meaningful FactSheets for AI services.

The remainder of the paper is organized as
follows. Section~\ref{sec-label} overviews related work, including
labeling, safety, and certification standards in other
industries. Section~\ref{sec-trust} provides more details on the key
issues to enable trust in AI systems. Section~\ref{sec-factsheet}
describes the AI service FactSheet in more detail, giving examples of
questions that it should include. 
In Section~\ref{sec-evolve}, we discuss how FactSheets can evolve from
a voluntary process to one that could be an industry
requirement. Section~\ref{sec-discuss} covers challenges,
opportunities, and future work needed to achieve the widespread usage
of AI service declarations of conformity. A proposed complete set of
sections and items for a FactSheet is included in the appendix, along with sample FactSheets for two exemplary fictitious services, fingerprint verification and trending topics in social media.

\section{Related Work}
\label{sec-label}

This section discusses related work in providing transparency in the
creation of AI services, as well as a brief survey of ensuring trust in non-AI systems.

\subsection{Transparency in AI}

Within the last year, several research groups have advocated standardizing and sharing information about training datasets and trained models. Gebru et al.\ proposed the use of \emph{datasheets for
datasets} as a way to expose and standardize information about public datasets, or datasets used in the development of commercial AI services and pre-trained models \cite{gebru-2018}. The datasheet would include provenance information, key characteristics, and relevant regulations, but also significant, yet more subjective information, such as potential bias, strengths and weaknesses, and suggested uses. Bender and Friedman propose a \emph{data statement schema}, as a way to capture and convey the information and properties of a dataset used in natural language processing (NLP) research and development \cite{data-statements}.  They argue that data statements should be included in most writing on NLP, including:  papers presenting new datasets, papers reporting experimental work with datasets, and documentation for NLP systems. 

Holland et al.\ outline the \emph{dataset nutrition label}, a diagnostic framework that provides a concise yet robust and standardized view of the core components of a dataset \cite{HollandHNJC2018}. Academic conferences such as the International AAAI Conference on Web and Social Media are also starting special tracks for dataset papers containing detailed descriptions, collection methods, and use cases.

Subsequent to the first posting of this paper \cite{factsheet-arxiv}, Mitchell et al.\ propose \emph{model cards} to convey information that characterizes the evaluation of a machine learning model in a variety of conditions and disclose the context in which models are intended to be used, details of the performance evaluation procedures, and other relevant information \cite{model-cards}. There is also budding activity on auditing and labeling algorithms for accuracy, bias, consistency, transparency, fairness and timeliness, in the industry \cite{ORCAA,SAFEAI}, but this audit does not cover several aspects of safety, security, and lineage.

Our proposal is distinguished from prior work in that we focus on the final AI service, a distinct concept from a single pre-trained machine learning model or dataset. Moreover, we take a broader view on trustworthy AI that extends beyond principles, values and ethical purpose to also include technical robustness and reliability \cite{EuropeanCommission2018}.

\subsection{Enabling Trust in Other Domains} 

Enabling trust in systems is not unique to AI.  This section
provides an overview of mechanisms used in other domains and
industries to achieve trust.  The goal is to understand existing
approaches to help inspire the right directions for enabling trust in
AI services.

\subsubsection{Standards Organizations}
Standardization organizations, such as the IEEE~\cite{ieee} and ISO~\cite{iso}, 
define standards along with the requirements that need to be satisifed for a product or a process to meet the standard. The product developer can self-report that a product meets the standard, though there are several cases, especially with ISO standards, where an independent accredited body will verify that the standards are met and provide the certification.

\subsubsection{Consumer Products}
The United States Consumer Product Safety Commission (CPSC) \cite{cpsc} requires a manufacturer or importer to declare its product as compliant with applicable consumer product safety requirements in a written or electronic declaration of conformity. In many cases, this can be self-reported by the manufacturer or importer, i.e.\ an SDoC.  However, in the case of children's products, it is mandatory to have the testing performed by a CPSC-accepted laboratory for compliance. Durable infant or toddler products must be marked with specialized tracking labels and must have a postage-paid customer registration card attached, to be used in case of a recall.

The National Parenting Center has a Seal of Approval program~\cite{seal-of-approval} that conducts testing on a variety of children's products involving interaction with the products by parents, children, and educators, who fill out questionnaires for the products they test. The quality of a product is determined based on factors like the product's level of desirability, sturdiness, and interactive stimulation. Both statistical averaging as well as comments from testers are examined before providing a Seal of Approval for the product.

\subsubsection{Finance}
In the financial industry, corporate bonds are rated by independent rating services~\cite{moodys,spratings} to help an investor assess the bond issuer's financial strength or its ability to pay a bond's principal and interest in a timely fashion. These letter-grade ratings range from AAA or Aaa for safe, `blue-chip' bonds to C or D for `junk' bonds. On the other hand, common-stock investments are not rated independently. Rather, the Securities and Exchange Commission (SEC) requires potential issuers of stock to submit specific registration documents that discloses extensive financial information about the company and risks associated with the future operations of the company. The SEC examines these documents, comments on them, and expects corrections based on the comments. The final product is a prospectus approved by the SEC that is available for potential buyers of the stock. 

\subsubsection{Software}
In the software area, there have been recent attempts to certify digital data repositories as `trusted.' Trustworthiness involves both the quality of the data and sustainable, reliable access to the data. The goal of certification is to enhance scientific reproducibility. The European Framework for Audit and Certification~\cite{trusted} has three levels of certification, Core, Extended, and Formal (or Bronze, Silver, and Gold), having different requirements, mainly to distinguish between the requirements of different types of data, e.g.\ research data vs.\ human health data vs.\ financial transaction data. The CoreTrustSeal~\cite{coretrustseal}, a private legal entity, provides a Bronze level certification to an interested data repository, for a nominal fee.

There have been several proposals in the literature for software certifications of various kinds. Ghosh and McGraw~\cite{ghosh} propose a certification process for testing software components for security properties. Their technique involves a process and a set of white-box and black-box testing procedures, that eventually results in a stamp of approval in the form of a digital signature. Schiller~\cite{schiller} proposes a certification process that starts with a checklist with yes/no answers provided by the developer, and determines which tests need to be performed on the software to certify it. Currit et al.~\cite{currit-1986}  describe a procedure for certifying the reliability of software before its release to the users. They predict the performance of the software on unseen inputs using the MTTF (mean time to failure) metric. Port and Wilf~\cite{port-2013} describe a procedure to certify the readiness for software release, understanding the tradeoff in cost of too early a release due to failures in the field, versus the cost in personnel and schedule delay arising from more extensive testing. Their technique involves the filling out of a questionnaire by the software developer called the Software Review and Certification Record (SRCR), which is `credentialed' with signatories who approve the document prior to the release decision. Heck et al.~\cite{heck-2010} also describe a software product certification model to certify legislative compliance or acceptability of software delivered during outsourcing. The basis for certification is a questionnaire to be filled out by the developer. The only acceptable answers to the questions are \emph{yes} and \emph{n/a} (not applicable). 

A different approach is taken in the CERT Secure Coding Standards~\cite{sei} of the Software Engineering Institute. Here the emphasis is on documenting best practices and coding standards for security purposes. The secure coding standards consist of guidelines about the types of security flaws that can be injected through development with specific programming languages. Each guideline offers precise information describing the cause and impact of violations, and examples of common non-compliant (flawed) and compliant (fixed) code. The organization also provides tools, which audits code to identify security flaws as indicated by violations of the CERT secure coding standards.

\subsubsection{Environmental Impact Statements}

Environment law in the United States requires that an environmental
impact statement (EIS) should be prepared prior to starting large
constructions.  An EIS is a document used as a tool for decision
making that describes positive and negative environmental effects of a
proposed action.  It is made available both to federal agencies and to
the public, and captures impacts to endangered species, air quality, water quality, cultural sites, and the socioeconomics of local communities.  The federal law, the National Environmental Policy Act, has inspired similar laws in various jurisdictions and in other fields beyond the environment.  Selbst has proposed an algorithmic impact statement for AI that follows the form and purpose of EISs \cite{Selbst2017}.

\subsubsection{Human Subjects}
In addition to products and technologies, another critical endeavor requiring trust is research involving human subjects. Institutional review boards (IRB) have precise reviewing protocols and requirements such as those presented in the Belmont Report \cite{Sims2010}. Items to be completed include statement of purpose, participant selection, procedures to be followed, harms and benefits to subjects, confidentiality, and consent documents. As AI services increasingly make inferences for people and about people~\cite{varshney-2015}, IRB requirements increasingly apply to them.

\subsubsection{Summary}
To ensure trust in products, industries have established a variety of practices to convey information about how a product is expected to perform when utilized by a consumer.  This information usually includes how the product was constructed and tested.  Some industries allow product creators to voluntarily provide this information, whereas others explicitly require it.  When the information is required, some industries require the information to be validated by a third party.  One would expect the latter scenario to occur in mature industries where there is confidence that the requirements strongly correlate with safety, reliability, and overall trust in the product. Mandatory external validation of nascent requirements in emerging industries may unnecessarily stifle the development of the industry.

\section{Elements of Trust in AI Systems}
\label{sec-trust}

We drive cars trusting the brakes will work when the pedal is pressed. We undergo laser eye surgery trusting the system to make the right decisions. We accept that the autopilot will operate an airplane, trusting that it will navigate correctly.  In all these cases, trust comes from confidence that the system will err extremely rarely, leveraging system training, exhaustive testing, experience, safety measures and standards, best practices, and consumer education. 

Every time new technology is introduced, it creates new challenges, safety issues, and potential hazards.  As the technology develops and matures, these issues are better understood, documented, and addressed. Human trust in technology is developed as users overcome perceptions of risk and uncertainty \cite{LiHV2008}, i.e., as they are able to assess the technology's performance, reliability, safety, and security.  Consumers do not yet trust AI like they trust other technologies because of inadequate attention given to the latter of these issues \cite{Scott2018}.  Making technical progress on safety and security is necessary but not sufficient to achieve trust in AI, however; the progress must be accompanied by the ability to measure and communicate the performance levels of the service on these dimensions in a standardized and transparent manner. One way to accomplish this is to provide such information via FactSheets for AI services.

Trust in AI services will come from: a) applying general safety and reliability engineering methodologies across the entire lifecycle of an AI service, b) identifying and addressing new, AI-specific issues and challenges in an ongoing and agile way, and c) creating standardized tests and transparent reporting mechanisms on how such a service operates and performs. In this section we outline several areas of concern and how they uniquely apply to AI.  The crux of this discussion is the manifestation of risk and uncertainty in machine learning, including that data distributions used for training are not always the ones that ideally should be used.  

\subsection{Basic Performance and Reliability} \label{sec-perf}

Statistical machine learning theory and practice is built around risk minimization. The particular loss function, whose expectation over the data distribution is considered to be the risk, depends on the task, e.g.\ zero-one loss for binary classification and mean squared error for regression. Different types of errors can be given different costs. Abstract loss functions may be informed by real-world quality metrics \cite{Wagstaff2012}, including context-dependent ones \cite{OlteanuTV2017}.  There is no particular standardization on the loss function, even broadly within application domains. Moreover, performance metrics that are not directly optimized are also often examined, e.g.\ area under the curve and normalized cumulative discounted gain.

The true expected value of the loss function can never be known and must be estimated empirically. There are several approaches and rules of thumb for estimating the risk,  but there is no standardization here either. Different groups make different choices (k-fold cross-validation, held-out samples, stratification, bootstrapping, etc.). Further notions of performance and reliability are the technical aspects of latency, throughput, and availability of the service, which are also not standardized for the specifics of AI workloads.

To develop trust in AI services from a basic performance perspective,
the choice of metrics and testing conditions should not be left to the
discretion of the supplier (who may choose conditions which present
the service in a favorable light), but should be codified and
standardized. The onerous requirement of third-party testing could be
avoided by ensuring that the specifications are precise, 
i.e., that
each metric is precisely defined to ensure consistency and enable
reproducibility  by AI service consumers.

For each metric a FactSheet should report the values under various
categories relevant to the expected consumers, (e.g.,
performance for various age groups, geographies, or genders) with the
goal of providing the right level of insight into the service, but
still preserving privacy.  We expect some metrics will be specific to
a domain, (e.g., finance, healthcare, manufacturing),
or a modality (e.g., visual, speech, text),  reflecting common 
practice of evaluation in that environment.

\subsection{Safety}

While typical machine learning performance metrics are measures of risk (the ones described in the previous section), we must also consider epistemic uncertainty when assessing the safety of a service \cite{Moller2012,VarshneyA2017}.  The main uncertainty in machine learning is an unknown mismatch between the training data distribution and the desired data distribution on which one would ideally train.  Usually that desired distribution is the true distribution encountered in operation (in this case the mismatch is known as dataset shift), but it could also be an idealized distribution that encodes preferred societal norms, policies, or regulations (imagine a more equitable world than what exists in reality).  One may map four general categories of strategies to achieve safety proposed in \cite{MollerH2008} to machine learning \cite{VarshneyA2017}: inherently safe design, safety reserves, safe fail, and procedural safeguards, all of which serve to reduce epistemic uncertainty. Interpretability of models is one example of inherently safe design.

\paragraph{Dataset Shift}

As the statistical relationship between features and labels changes
over time, known as dataset shift, the mismatch between the training
distribution and the distribution from which test samples are being
drawn increases. A well-known reason for performance degradation,
dataset shif is a common cause of frustration and loss of trust for AI
service consumers.  Dataset shift can be detected and corrected using a multitude of methods \cite{GamaZBPB2014}.  The sensitivity of performance of different models to dataset shift varies and should be part of a testing protocol. To the best of our knowledge, there does not yet exist any standard for how to conduct such testing.
To mitigate this risk a FactSheet should contain demographic
information about the training and test datasets that report the
various outcomes for each group of interest as specified in
Section~\ref{sec-perf}. 

\paragraph{Fairness}

AI fairness is a rapidly growing topic of inquiry
\cite{HajianBC2016}. There are many different definitions of fairness
(some of which provably conflict) that are appropriate in varying
contexts.  The concept of fairness relies on protected attributes
(also context-dependent) such as race, gender, caste, and
religion. For fairness, we insist on some risk measure being
approximately equal in groups defined by the protected
attributes. Unwanted biases in training data, due to either prejudice
in labels or under-/over-sampling, lead to unfairness and can be
checked using statistical tests on datasets or models
\cite{BarocasS2016,aif360-oct-2018}. One can think of bias as the
mismatch between the training data distribution and a desired fair
distribution. Applications such as lending have legal requirements on
fairness in decision making, e.g.\ the Equal Credit Opportunity Act in
the United States. Although the parity definitions and computations in
such applications are explicit,  the interpretation of the numbers is
subjective: there are no immutable thresholds on fairness metrics (e.g.,
the well-known 80\% rule~\cite{FeldmanFMSV2015}) that are aplied in isolation of context. 

\paragraph{Explainability}

Directly interpretable machine learning (in contrast to post hoc interpretation) \cite{Rudin2014}, in which a person can look at a model and understand what it does, reduces epistemic uncertainty and increases safety because quirks and vagaries of training dataset distributions that will not be present in distributions during deployment can be identified by inspection \cite{VarshneyA2017}. Different users have different needs from explanations, and there is not yet any satisfactory quantitative definition of interpretability (and there may never be) \cite{DoshiVelezK2017}.  Recent regulations in the European Union require `meaningful' explanations, but it is not clear what constitutes a meaningful explanation. 

\subsection{Security}

AI services can be attacked by adversaries in various ways \cite{PapernotMJFCS2016}. Small imperceptible perturbations could cause AI services to misclassify inputs to any label that attackers desire; training data and models can be poisoned, allowing attackers to worsen performance (similar to concept drift but deliberate); and sensitive information about data and models can be stolen by observing the outputs of a service for different inputs. Services may be instrumented to detect such attacks and may also be designed with defenses \cite{NicolaeSTRWZBCLME2018}. New research proposes certifications for defenses against adversarial examples \cite{RaghunathanSL2018}, but these are not yet practical. 

\subsection{Lineage}

Once performance, safety, and security are sufficient to engender trust, we must also ensure that we track and maintain the provenance of datasets, metadata, models along with their hyperparameters, and test results. Users, those potentially affected, and third parties, such as regulators, must be able to audit the systems underlying the services. Appropriate parties may need the ability to reproduce past outputs and track outcomes. Specifically, one should be able to determine the exact version of the service deployed at any point of time in the past, how many times the service was retrained and associated details like hyperparameters used for each training episode, training dataset used, how accuracy and safety metrics have evolved over time, the feedback data received by the service, and the triggers for retraining and improvement.  This information may span multiple organizations when a service is built by multiple parties.

\section{Items in a FactSheet} 
\label{sec-factsheet}

In this section we provide an overview of the items that should be
addressed in a FactSheet. See the appendix for the complete list of
items. To illustrate how these items might be completed in practice,
we also include two sample FactSheets in the appendix: one for a fictitious fingerprint verification service and one for a trending topics service. 

The items are grouped into several categories aligned with the elements of trust. The categories are: statement of purpose, basic performance, safety, security, and lineage.  They cover various aspects of service development, testing, deployment and maintenance: from information about the data the service is trained on, to underlying algorithms, test setup, test results, and performance benchmarks, to the way the service is maintained and retrained (including automatic adaptation).
    
The items are devised to aid the user in understanding how the service works, in determining if the service is appropriate for the intended application, and in comprehending its strengths and limitations. The identified items are not intended to be definitive. If a question is not applicable to a given service, it can simply be ignored. In some cases, the service supplier may not wish to disclose details of the service for competitive reasons. For example, a supplier of a commercial fraud detection service for healthcare insurance claims may choose not to reveal the details of the underlying algorithm; nevertheless, the supplier should be able to indicate the class of algorithm used, provide sample outputs along with explanations of the algorithmic decisions leading to the outputs. More consequential applications will likely require more comprehensive completion of items.

A few examples of items a FactSheet might include are:
\begin{itemize}
\item What is the intended use of the service output?
\item What algorithms or techniques does this service implement?
\item Which datasets was the service tested on? (Provide links to datasets that were used for testing, along with corresponding datasheets.)
\item Describe the testing methodology.
\item Describe the test results.
\item Are you aware of possible examples of bias, ethical issues, or other safety risks as a result of using the service?
\item Are the service outputs explainable and/or interpretable?
\item For each dataset used by the service: Was the dataset checked for bias? What efforts were made to ensure that it is fair and representative?
\item Does the service implement and perform any bias detection and remediation?
\item What is the expected performance on unseen data or data with different distributions?
\item Was the service checked for robustness against adversarial attacks?
\item When were the models last updated?
\end{itemize}

As such a declaration is refined, and testing procedures for performance, robustness to concept drift, explainability, and robustness to attacks are further codified, the FactSheet may refer to standardized test protocols instead of providing descriptive details.

Since completing a FactSheet can be laborious, 
we expect most of the information to be populated as part
of the AI service creation process in a secure auditable manner.
A FactSheet will be created once and associated with a service, but 
can continually be augmented, without removing previous information,
i.e., results are added from more tests, but results cannot be removed.
Any changes made to the service will prompt the creation of a new
version of the FactSheet for the new model.  Thus, these FactSheets will be
treated as a series of immutable artifacts.

This information can be used to more accurately monitor a deployed service by
comparing deployed metrics with those that were seen during
development and taking appropriate action when unexpected behavior is
detected.

\section{The Evolution of FactSheet Adoption}
\label{sec-evolve}
We expect that AI will soon go through the same evolution that other technologies have gone through (cf.\ \cite{gebru-2018} for an excellent review of the evolution of safety standards in different industries).  We propose that FactSheets be initially voluntary for several reasons. First, discussion and feedback from  multiple parties representing suppliers and consumers of AI services is needed to determine the final set of items and format of FactSheets. So, an initial voluntary period to allow this discussion to  occur is needed. Second, there needs to be a balance between the needs of AI service consumers with the freedom to innovate for AI service producers. Although producing a FactSheet will initially be an additional burden to an AI service producer, we expect market feedback from AI service consumers to encourage this creation.

Because of peer pressure to conform \cite{ben-AaronDDW2017}, FactSheets could become a de facto requirement similar to Energy Star labeling of the energy efficiency of appliances. They will serve to reduce information asymmetry between supplier and consumer, where consumers are currently unaware of important properties of a service, such as its intended use, its performance metrics, and information about fairness, explainability, safety, and security. In particular, consumers in many businesses do not have the requisite expertise to evaluate various AI services available in the marketplace; uninformed or incorrect choices can result in suboptimal business performance. By creating easily consumable FactSheets, suppliers can accrue a competitive advantage by capturing consumers' trust. Moreover, with such transparency, FactSheets should serve to allow better functioning of AI service marketplaces and prevent a so-called `market for lemons' \cite{Akerlof1970}.  A counter-argument to voluntary compliance and self-regulation argues that while participation of industry is welcome, this should not stand in the way of legislation and governmental regulation  \cite{nemitz2018constitutional}.

FactSheet adoption could potentially lead to an eventual system of third-party certification \cite{SrivastavaR2018}, but probably only for services catering to applications with the very highest of stakes, to regulated business processes and enterprise applications, and to applications originating in the public sector \cite{McAllister2014,ANSI}. Children's toys are an example category of consumer products in which an SDoC is not enough and certification is required. If an AI service is already touching on a regulation from a specific industry in which it is being used, its FactSheet will serve as a tool for better compliance.

\section{Discussion and Future Work} 
\label{sec-discuss}

One may wonder why AI should be held to a higher standard (FactSheets) than non-AI software and services in the same domain.  Non-AI software include several artifacts beyond the code, such as design documents, program flow charts, and test plans that can provide transparency to concerned consumers.  Since AI services do not contain any of these, and the generated code may not be easily understandable, there is a higher demand to enhance transparency through FactSheets.

Although FactSheets enable AI services producers to provide information about the intent and construction of their service so that educated consumers can make informed decisions, consumers may still, innocently or maliciously, use the service for purposes other than those intended. FactSheets cannot fully protect against such use, but can form the basis of service level agreements.

Some components of an AI service may be produced by organizations other than the service supplier. For example, the dataset may be obtained from a third party, or the service may be a composition of models, some of which are produced by another organization. In such cases, the FactSheet for the composed service would need to include information from the supplying organizations. Ideally, those organizations would produce FactSheets for their components, enabling the composing organization to provide a complete FactSheet. This complete FactSheet could include the component FactSheets along with any necessary additional information. In some cases, the demands for transparency on the composing organization may be greater than on the component organization; market forces will require the component organization to provide more transparency to retain their relation with the composing organization. This is analogous to other industries, like retail, where retailers push demands on their suppliers to meet the expectations of the retailers' customers. In these situations the provenance of the information among organizations will need to be tracked.

\section{Summary and Conclusion} 
\label{sec-conclusion}

In this paper, we continue in the research direction established by datasheets or nutrition labels for datasets to examine trusted AI at the functional level rather than at the component level. We discuss the several elements of AI services that are needed for people to trust them, including task performance, safety, security, and maintenance of lineage.  The final piece to build trust is transparent documentation about the service, which we see as a variation on declarations of conformity for consumer products.  We propose a starting point to a voluntary AI service supplier's declaration of conformity.  Further discussion among multiple parties is required to standardize protocols for testing AI services and determine the final set of items and format that AI service FactSheets will take.

We envision that suppliers will voluntarily populate and release FactSheets for their services to remain competitive in the market. The evolution of the marketplace of AI services may eventually lead to an ecosystem of third party testing and verification laboratories, services, and tools. We also envision the automation of nearly the entire FactSheet as part of the build and runtime environments of AI services. Moreover, it is not difficult to imagine FactSheets being automatically posted to distributed, immutable ledgers such as those enabled by blockchain technologies.

We see our work as a first step at defining which questions to ask and metrics to measure towards development and adoption of broader industry practices and standards. We see a parallel between the issue of trusted AI today and the rise of digital certification during the Internet revolution. The digital certification market `bootstrapped' the Internet, ushering in a new era of `transactions' such as online banking and benefits enrollment that we take for granted today. In a similar vein, we can see AI service FactSheets ushering in a new era of trusted AI end points and bootstrapping broader adoption.

\bibliographystyle{IEEETran}
\bibliography{sdocs}

\appendix

\section{Proposed FactSheet Items}
Below we list example questions that a FactSheet for an AI service could include. The set of questions we provide here is not intended to be definitive, but rather to open a conversation about what aspects should be covered. 

To illustrate how these questions could be answered, we provide two examples for fictitious AI services: a fingerprint verification service (Appendix~\ref{fingerprint}) and a trending topics service (Appendix \ref{trending}). 
However, given that the examples we provide are fictitious, we would
expect an actual service provider to answer these questions in much
more detail. For instance, they would be able to better characterize
an API that actually exists. Our example answers are mainly to provide
additional insight about the type of information we would find in a FactSheet. 

\subsection*{Statement of purpose}

The following questions are aimed at providing an overview of the
service provider and of the intended uses for the service. Valid
answers include ``N/A'' (not applicable) and ``Proprietary'' (cannot be publicly disclosed, usually for competitive reasons).

\vspace{1.2em} \noindent \textbf{General}

\begin{itemize}
\item Who are ``you'' (the supplier) and what type of services do you typically offer (beyond this particular service)?

\item What is this service about? 

\vspace{0.6em}
\renewcommand{\arraystretch}{0.4}
{\small \begin{tabular}{|p{6.9cm}|}
\hline 
\begin{itemize}
\item Briefly describe the service.
\item When was the service first released? When was the last release?
\item Who is the target user?
\end{itemize} 
\vspace{-0.9em}
\\\hline
\end{tabular}}
\vspace{0.05em}

\item Describe the outputs of the service.

\item What algorithms or techniques does this service implement?

\vspace{0.6em}
\renewcommand{\arraystretch}{0.4}
{\small \begin{tabular}{|p{6.9cm}|}
\hline
\begin{itemize}
\item Provide links to technical papers.
\end{itemize}
\vspace{-0.9em}
\\\hline
\end{tabular}}
\vspace{0.05em}

\item What are the characteristics of the development team?

\vspace{0.6em}
\renewcommand{\arraystretch}{0.4}
{\small \begin{tabular}{|p{6.9cm}|}
\hline
\begin{itemize}
\item Do the teams charged with developing and maintaining this service reflect a diversity of opinions, backgrounds, and thought?
\end{itemize}
\vspace{-0.9em}
\\\hline
\end{tabular}}
\vspace{0.05em}

\item Have you updated this FactSheet before?

\vspace{0.6em}
\renewcommand{\arraystretch}{0.4}
{\small \begin{tabular}{|p{6.9cm}|}
\hline
\begin{itemize}
\item When and how often?
\item What sections have changed?
\item Is the FactSheet updated every time the service is retrained or updated?
\end{itemize}
\vspace{-0.9em}
\\\hline
\end{tabular}}
\vspace{0.05em}

\end{itemize}

\noindent \textbf{Usage}

\begin{itemize}
\item What is the intended use of the service output?

\vspace{0.6em}
\renewcommand{\arraystretch}{0.4}
{\small \begin{tabular}{|p{6.9cm}|}
\hline
\begin{itemize}
\item Briefly describe a simple use-case.
\end{itemize}
\vspace{-0.9em}
\\\hline
\end{tabular}}
\vspace{0.05em}

\item What are the key procedures followed while using the service?

\vspace{0.6em}
\renewcommand{\arraystretch}{0.4}
{\small \begin{tabular}{|p{6.9cm}|}
\hline
\begin{itemize}
\item How is the input provided? By whom?
\item How is the output returned?
\end{itemize}
\vspace{-0.9em}
\\\hline
\end{tabular}}
\vspace{0.05em}

\end{itemize}

\noindent \textbf{Domains and applications}
\begin{itemize}
\item What are the domains and applications the service was tested on or used for?

\vspace{0.6em}
\renewcommand{\arraystretch}{0.4}
{\small \begin{tabular}{|p{6.9cm}|}
\hline
\begin{itemize}
\item Were domain experts involved in the development, testing, and deployment? Please elaborate.
\end{itemize}
\vspace{-0.9em}
\\\hline
\end{tabular}}
\vspace{0.05em}

\item How is the service being used by your customers or users?

\vspace{0.6em}
\renewcommand{\arraystretch}{0.4}
{\small \begin{tabular}{|p{6.9cm}|}
\hline
\begin{itemize}
\item Are you enabling others to build a solution by providing a cloud service or is your application end-user facing?
\item Is the service output used as-is, is it fed directly into another tool or actuator, or is there human input/oversight before use?
\item Do users rely on pre-trained/canned models or can they train their own models?
\item Do your customers typically use your service in a time critical setup (e.g. they have limited time to evaluate the output)? Or do they incorporate it in a slower decision making process? Please elaborate.
\end{itemize}
\vspace{-0.9em}
\\\hline
\end{tabular}}
\vspace{0.05em}

\item List applications that the service has been used for in the past.

\vspace{0.6em}
\renewcommand{\arraystretch}{0.4}
{\small \begin{tabular}{|p{6.9cm}|}
\hline
\begin{itemize}
\item Please provide information about these applications or relevant pointers.
\item Please provide key performance results for those applications.
\end{itemize}
\vspace{-0.9em}
\\\hline
\end{tabular}}
\vspace{0.05em}

\item Other comments?
\end{itemize}

\subsection*{Basic Performance}
The following questions aim to offer an overall assessment of the service performance.

\vspace{1em} \noindent \textbf{Testing by service provider}

\begin{itemize}
\item Which datasets was the service tested on? (e.g., links to datasets that were used for testing, along with corresponding datasheets)

\vspace{0.6em}
\renewcommand{\arraystretch}{0.4}
{\small \begin{tabular}{|p{6.9cm}|}
\hline

\begin{itemize}
\item List the test datasets and provide links to these datasets.
\item Do the datasets have an associated datasheet? If yes, please attach.
\item Could these datasets be used for independent testing of the service? Did the data need to be changed or sampled before use? 
\end{itemize}
\vspace{-0.9em}
\\\hline
\end{tabular}}
\vspace{0.05em}

\item Describe the testing methodology.

\vspace{0.6em}
\renewcommand{\arraystretch}{0.4}
{\small \begin{tabular}{|p{6.9cm}|}
\hline

\begin{itemize}
\item Please provide details on train, test and holdout data.
\item What performance metrics were used? (e.g. accuracy, error rates, AUC, precision/recall)
\item Please briefly justify the choice of metrics.
\end{itemize}
\vspace{-0.9em}
\\\hline
\end{tabular}}
\vspace{0.05em}

\item Describe the test results.

\vspace{0.6em}
\renewcommand{\arraystretch}{0.4}
{\small \begin{tabular}{|p{6.9cm}|}
\hline

\begin{itemize}
\item Were latency, throughput, and availability measured?
\item If yes, briefly include those metrics as well.
\end{itemize}
\vspace{-0.9em}
\\\hline
\end{tabular}}
\vspace{0.05em}

\end{itemize}

\vspace{1em} \noindent \textbf{Testing by third parties}

\begin{itemize}
\item Is there a way to verify the performance metrics (e.g., via a service API )?

\vspace{0.6em}
\renewcommand{\arraystretch}{0.4}
{\small \begin{tabular}{|p{6.9cm}|}
\hline

\begin{itemize}
\item Briefly describe how a third party could independently verify the performance of the service.
\item Are there benchmarks publicly available and adequate for testing the service.
\end{itemize}
\vspace{-0.9em}
\\\hline
\end{tabular}}
\vspace{0.05em}

\item In addition to the service provider, was this service tested by any third party?

\vspace{0.6em}
\renewcommand{\arraystretch}{0.4}
{\small \begin{tabular}{|p{6.9cm}|}
\hline

\begin{itemize}
\item Please list all third parties that performed the testing.
\item Also, please include information about the tests and test results.
\end{itemize}
\vspace{-0.9em}
\\\hline
\end{tabular}}
\vspace{0.05em}

\item Other comments?

\end{itemize}

\subsection*{Safety}

The following questions aim to offer insights about potential unintentional harms, and mitigation efforts to eliminate or minimize those harms.

\vspace{1em} \noindent \textbf{General}

\begin{itemize}
\item Are you aware of possible examples of bias, ethical issues, or other safety risks as a result of using the service?

\vspace{0.6em}
\renewcommand{\arraystretch}{0.4}
{\small \begin{tabular}{|p{6.9cm}|}
\hline

\begin{itemize}
\item Were the possible sources of bias or unfairness analyzed?
\item Where do they arise from: the data? the particular techniques being implemented? other sources?
\item Is there any mechanism for redress if individuals are negatively affected?
\end{itemize}
\vspace{-0.9em}
\\\hline
\end{tabular}}
\vspace{0.05em}

\item Do you use data from or make inferences about individuals or groups of individuals. Have you obtained their consent?

\vspace{0.6em}
\renewcommand{\arraystretch}{0.4}
{\small \begin{tabular}{|p{6.9cm}|}
\hline

\begin{itemize}
\item How was it decided whose data to use or about whom to make inferences?
\item Do these individuals know that their data is being used or that inferences are being made about them? What were they told? When were they made aware? What kind of consent was needed from them? What were the procedures for gathering consent? Please attach the consent form to this declaration.
\item What are the potential risks to these individuals or groups? Might the service output interfere with individual rights? How are these risks being handled or minimized?
\item What trade-offs were made between the rights of these individuals and business interests?
\item Do they have the option to withdraw their data?  Can they opt out from inferences being made about them? What is the withdrawal procedure?
\end{itemize}
\vspace{-0.9em}
\\\hline
\end{tabular}}
\vspace{0.05em}

\end{itemize}

\noindent \textbf{Explainability}

\begin{itemize}
\item Are the service outputs explainable and/or interpretable?

\vspace{0.6em}
\renewcommand{\arraystretch}{0.4}
{\small \begin{tabular}{|p{6.9cm}|}
\hline

\begin{itemize}
\item Please explain how explainability is achieved (e.g. directly explainable algorithm, local explainability, explanations via examples).
\item Who is the target user of the explanation (ML expert, domain expert, general consumer, etc.)
\item Please describe any human validation of the explainability of the algorithms
\end{itemize}
\vspace{-0.9em}
\\\hline
\end{tabular}}
\vspace{0.05em}

\end{itemize}

\vspace{1.2em} \noindent \textbf{Fairness}

\begin{itemize}
\item For each dataset used by the service: Was the dataset checked for bias? What efforts were made to ensure that it is fair and representative?

\vspace{0.6em}
\renewcommand{\arraystretch}{0.4}
{\small \begin{tabular}{|p{6.9cm}|}
\hline

\begin{itemize}
\item Please describe the data bias policies that were checked (such as with respect to known protected attributes), bias checking methods, and results (e.g., disparate error rates across different groups).
\item Was there any bias remediation performed on this dataset? Please provide details about the value of any bias estimates before and after it.
\item What techniques were used to perform the remediation? Please provide links to relevant technical papers.
\item How did the value of other performance metrics change as a result?
\end{itemize}
\vspace{-0.9em}
\\\hline
\end{tabular}}
\vspace{0.05em}

\item Does the service implement and perform any bias detection and remediation?

\vspace{0.6em}
\renewcommand{\arraystretch}{0.4}
{\small \begin{tabular}{|p{6.9cm}|}
\hline

\begin{itemize} 
\item Please describe model bias policies that were checked, bias checking methods, and results (e.g., disparate error rates across different groups).
\item What procedures were used to perform the remediation? Please provide links or references to corresponding technical papers.
\item Please provide details about the value of any bias estimates before and after such remediation.
\item How did the value of other performance metrics change as a result?
\end{itemize}
\vspace{-0.9em}
\\\hline
\end{tabular}}
\vspace{0.05em}

\end{itemize}

\noindent \textbf{Concept Drift}

\begin{itemize}

\item What is the expected performance on unseen data or data with different distributions?

\vspace{0.6em}
\renewcommand{\arraystretch}{0.4}
{\small \begin{tabular}{|p{6.9cm}|}
\hline

\begin{itemize}
\item Please describe any relevant testing done along with test results.
\end{itemize}

\vspace{-0.9em}
\\\hline
\end{tabular}}
\vspace{0.05em}

\item Does your system make updates to its behavior based on newly ingested data?

\vspace{0.6em}
\renewcommand{\arraystretch}{0.4}
{\small \begin{tabular}{|p{6.9cm}|}
\hline

\begin{itemize}
\item Is the new data uploaded by your users? Is it generated by an automated process? Are the patterns in the data largely static or do they change over time?
\item Are there any performance guarantees/bounds?
\item Does the service have an automatic feedback/retraining loop, or is there a human in the loop?
\end{itemize}
\vspace{-0.9em}
\\\hline
\end{tabular}}
\vspace{0.05em}

\item How is the service tested and monitored for model or performance drift over time?

\vspace{0.6em}
\renewcommand{\arraystretch}{0.4}
{\small \begin{tabular}{|p{6.9cm}|}
\hline

\begin{itemize}
\item If applicable, describe any relevant testing along with test results.
\end{itemize}

\vspace{-0.9em}
\\\hline
\end{tabular}}
\vspace{0.05em}

\item How can the service be checked for correct, expected output when new data is added?

\item Does the service allow for checking for differences between training and usage data?

\vspace{0.6em}
\renewcommand{\arraystretch}{0.4}
{\small \begin{tabular}{|p{6.9cm}|}
\hline

\begin{itemize}
\item Does it deploy mechanisms to alert the user of the difference?
\end {itemize}

\vspace{-0.9em}
\\\hline
\end{tabular}}
\vspace{0.05em}

\item Do you test the service periodically?

\vspace{0.6em}
\renewcommand{\arraystretch}{0.4}
{\small \begin{tabular}{|p{6.9cm}|}
\hline

\begin{itemize}
\item Does the testing includes bias or fairness related aspects?
\item How has the value of the tested metrics evolved over time?
\end{itemize}

\vspace{-0.9em}
\\\hline
\end{tabular}}
\vspace{0.05em}

\item Other comments?

\end{itemize}

\subsection*{Security}

The following questions aim to assess the susceptibility to deliberate harms such as attacks by adversaries.

\begin{itemize}
\item How could this service be attacked or abused? Please describe.

\item List applications or scenarios for which the service is not suitable.

\vspace{0.6em}
\renewcommand{\arraystretch}{0.4}
{\small \begin{tabular}{|p{6.9cm}|}
\hline

\begin{itemize}
\item Describe specific concerns and sensitive use cases.
\item Are there any procedures in place to ensure that the service will not be used for these applications?
\end{itemize}

\vspace{-0.9em}
\\\hline
\end{tabular}}
\vspace{0.05em}

\item How are you securing user or usage data?

\vspace{0.6em}
\renewcommand{\arraystretch}{0.4}
{\small \begin{tabular}{|p{6.9cm}|}
\hline

\begin{itemize}
\item Is usage data from service operations retained and stored?
\item How is the data being stored? For how long is the data stored?
\item Is user or usage data being shared outside the service? Who has access to the data?
\end{itemize}

\vspace{-0.9em}
\\\hline
\end{tabular}}
\vspace{0.05em}

\item Was the service checked for robustness against adversarial attacks?

\vspace{0.6em}
\renewcommand{\arraystretch}{0.4}
{\small \begin{tabular}{|p{6.9cm}|}
\hline

\begin{itemize}
\item Describe robustness policies that were checked, the type of attacks considered, checking methods, and results.
\end{itemize}

\vspace{-0.9em}
\\\hline
\end{tabular}}
\vspace{0.05em}

\item What is the plan to handle any potential security breaches?

\vspace{0.6em}
\renewcommand{\arraystretch}{0.4}
{\small \begin{tabular}{|p{6.9cm}|}
\hline

\begin{itemize}
\item Describe any protocol that is in place.
\end{itemize}

\vspace{-0.9em}
\\\hline
\end{tabular}}
\vspace{0.05em}

\item Other comments?
\end{itemize}

\subsection*{Lineage}

The following questions aim to overview how the service provider keeps track of details that might be required in the event of an audit by a third party, such as in the case of harm or suspicion of harm.

\vspace{1em} \noindent \textbf{Training Data}
\begin{itemize}
\item Does the service provide an as-is/canned model? Which datasets was the service trained on?

\vspace{0.6em}
\renewcommand{\arraystretch}{0.4}
{\small \begin{tabular}{|p{6.9cm}|}
\hline

\begin{itemize}
\item List the training datasets.
\item Where there any quality assurance processes employed while the data was collected or before use?
\item Were the datasets used for training built-for-purpose or were they re-purposed/adapted? Were the datasets created specifically for the purpose of training the models offered by this service?
\end{itemize}

\vspace{-0.9em}
\\\hline
\end{tabular}}
\vspace{0.05em}

\item For each dataset: Are the training datasets publicly available?

\vspace{0.6em}
\renewcommand{\arraystretch}{0.4}
{\small \begin{tabular}{|p{6.9cm}|}
\hline

\begin{itemize}
\item Please provide a link to the datasets or the source of the datasets.
\end{itemize}

\vspace{-0.9em}
\\\hline
\end{tabular}}
\vspace{0.05em}

\item For each dataset: Does the dataset have a datasheet or data statement?

\vspace{0.6em}
\renewcommand{\arraystretch}{0.4}
{\small \begin{tabular}{|p{6.9cm}|}
\hline

\begin{itemize} 
\item If available, attach the datasheet; otherwise, provide answers to questions from the datasheet as appropriate [to insert citation]
\end{itemize}

\vspace{-0.9em}
\\\hline
\end{tabular}}
\vspace{0.05em}

\item Did the service require any transformation of the data in addition to those provided in the datasheet?

\item Do you use synthetic data?

\vspace{0.6em}
\renewcommand{\arraystretch}{0.4}
{\small \begin{tabular}{|p{6.9cm}|}
\hline

\begin{itemize}
\item When? How was it created?
\item Briefly describe its properties and the creation procedure.

\end{itemize}

\vspace{-0.9em}
\\\hline
\end{tabular}}
\vspace{0.05em}

\end{itemize}

\noindent \textbf{Trained Models}
\begin{itemize}
\item How were the models trained?

\vspace{0.6em}
\renewcommand{\arraystretch}{0.4}
{\small \begin{tabular}{|p{6.9cm}|}
\hline

\begin{itemize}
\item Please provide specific details (e.g., hyperparameters).
\end{itemize}

\vspace{-0.9em}
\\\hline
\end{tabular}}
\vspace{0.05em}

\item When were the models last updated?

\vspace{0.6em}
\renewcommand{\arraystretch}{0.4}
{\small \begin{tabular}{|p{6.9cm}|}
\hline

\begin{itemize}
\item How much did the performance change with each update?
\item How often are the models retrained or updated?
\end{itemize}

\vspace{-0.9em}
\\\hline
\end{tabular}}
\vspace{0.05em}

\item Did you use any prior knowledge or re-weight the data in any way before training?

\item Other comments?
\end{itemize}

\clearpage

\section{Sample FactSheet for a Fingerprint Verification Service}
\label{fingerprint}

\subsection*{Statement of purpose}

The following questions are aimed at providing an overview of the service provider and of the intended uses for the service. Valid answers include ``N/A'' (not applicable) and ``Proprietary'' (cannot be publicly disclosed, usually for competitive reasons).

\vspace{1.2em} \noindent \textbf{General}

\begin{itemize}
\item Who are ``you'' (the supplier) and what type of services do you typically offer (beyond this particular service)?

\vspace{0.6em}{\color{blue}Raj Kumar Biometrics Services, Ltd. The only service we offer at present is fingerprint verification.}\vspace{0.6em}

\item What is this service about? 

\vspace{0.6em}
\renewcommand{\arraystretch}{0.4}
{\small \begin{tabular}{|p{6.9cm}|}
\hline 
\begin{itemize}
\item Briefly describe the service.
\item When was the service first released? When was the last release?
\item Who is the target user?
\end{itemize} 
\vspace{-0.9em}
\\\hline
\end{tabular}}
\vspace{0.05em}

\vspace{0.6em}{\color{blue}The service takes an ordered pair of fingerprint image and identity and returns a 1 if the image matches the image corresponding to that identity in the database. The service accepts 500 dpi images acquired using optical sensors. The v1.0 algorithm was created on June 30, 2005. The current algorithm v1.7 was created on April 12, 2010. The algorithm was released as a cloud service on August 10, 2017. The target user is a manufacturer who creates physical access control systems as well as other entities interested in physical or informational access control.}\vspace{0.6em}

\item Describe the outputs of the service.

\vspace{0.6em}{\color{blue}A binary verification label.}\vspace{0.6em}

\item What algorithms or techniques does this service implement?

\vspace{0.6em}
\renewcommand{\arraystretch}{0.4}
{\small \begin{tabular}{|p{6.9cm}|}
\hline
\begin{itemize}
\item Provide links to technical papers.
\end{itemize}
\vspace{-0.9em}
\\\hline
\end{tabular}}
\vspace{0.05em}

\vspace{0.6em}{\color{blue}P. Baldi and Y. Chauvin, ``Neural networks for fingerprint recognition,'' Neural Computation, vol. 5, no. 3, pp. 402--418, 1993.}\vspace{0.6em}

\item What are the characteristics of the development team?

\vspace{0.6em}
\renewcommand{\arraystretch}{0.4}
{\small \begin{tabular}{|p{6.9cm}|}
\hline
\begin{itemize}
\item Do the teams charged with developing and maintaining this service reflect a diversity of opinions, backgrounds, and thought?
\end{itemize}
\vspace{-0.9em}
\\\hline
\end{tabular}}
\vspace{0.05em}

\vspace{0.6em}{\color{blue}The service was developed by 3 graduates of Delhi College of Engineering. It was made into a cloud service and is maintained by 2 graduates of Amity University.}\vspace{0.6em}

\item Have you updated this FactSheet before?

\vspace{0.6em}
\renewcommand{\arraystretch}{0.4}
{\small \begin{tabular}{|p{6.9cm}|}
\hline
\begin{itemize}
\item When and how often?
\item What sections have changed?
\item Is the FactSheet updated every time the service is retrained or updated?
\end{itemize}
\vspace{-0.9em}
\\\hline
\end{tabular}}
\vspace{0.05em}

\vspace{0.6em}{\color{blue}This is our first release of FactSheet. We plan to release a new FactSheet when we release v1.8.}\vspace{0.6em}
\end{itemize}

\noindent \textbf{Usage}

\begin{itemize}
\item What is the intended use of the service output?

\vspace{0.6em}
\renewcommand{\arraystretch}{0.4}
{\small \begin{tabular}{|p{6.9cm}|}
\hline
\begin{itemize}
\item Briefly describe a simple use-case.
\end{itemize}
\vspace{-0.9em}
\\\hline
\end{tabular}}
\vspace{0.05em}

\vspace{0.6em}{\color{blue}A locks manufacturer is creating a biometrics-driven access control system that it will sell to call centers. This internet-enabled system will include an optical sensor for fingerprints and a keypad for the user to enter a 7 digit identification number. The system will acquire the fingerprint and identification number and transmit them to our service via a RESTful API. If the image matches the image for that user in the previously acquired database, it will return a positive result and the system will unlock.}\vspace{0.6em}

\item What are the key procedures followed while using the service?

\vspace{0.6em}
\renewcommand{\arraystretch}{0.4}
{\small \begin{tabular}{|p{6.9cm}|}
\hline
\begin{itemize}
\item How is the input provided? By whom?
\item How is the output returned?
\end{itemize}
\vspace{-0.9em}
\\\hline
\end{tabular}}
\vspace{0.05em}

\vspace{0.6em}{\color{blue}The end user supplies his or her fingerprint via an optical sensor which digitizes it and transmits it to the service. The binary output is returned to the access-control system.}\vspace{0.6em}

\end{itemize}

\noindent \textbf{Domains and applications}
\begin{itemize}
\item What are the domains and applications the service was tested on or used for?

\vspace{0.6em}
\renewcommand{\arraystretch}{0.4}
{\small \begin{tabular}{|p{6.9cm}|}
\hline
\begin{itemize}
\item Were domain experts involved in the development, testing, and deployment? Please elaborate.
\end{itemize}
\vspace{-0.9em}
\\\hline
\end{tabular}}
\vspace{0.05em}

\vspace{0.6em}{\color{blue}The service has only been tested on fingerprint verification of call center employees with no domain experts involved.}\vspace{0.6em}

\item How is the service being used by your customers or users?

\vspace{0.6em}
\renewcommand{\arraystretch}{0.4}
{\small \begin{tabular}{|p{6.9cm}|}
\hline
\begin{itemize}
\item Are you enabling others to build a solution by providing a cloud service or is your application end-user facing?
\item Is the service output used as-is, is it fed directly into another tool or actuator, or is there human input/oversight before use?
\item Do users rely on pre-trained/canned models or can they train their own models?
\item Do your customers typically use your service in a time critical setup (e.g. they have limited time to evaluate the output)? Or do they incorporate it in a slower decision making process? Please elaborate.
\end{itemize}
\vspace{-0.9em}
\\\hline
\end{tabular}}
\vspace{0.05em}

\vspace{0.6em}{\color{blue}Our service is a cloud service for access control system manufacturers. The output is directly fed into an actuator. Users can only rely on pre-trained models, but will necessarily upload a database of individual identifiers with their fingerprints. The service requires outputs be given with small delay.}\vspace{0.6em}

\item List applications that the service has been used for in the past.

\vspace{0.6em}
\renewcommand{\arraystretch}{0.4}
{\small \begin{tabular}{|p{6.9cm}|}
\hline
\begin{itemize}
\item Please provide information about these applications or relevant pointers.
\item Please provide key performance results for those applications.
\end{itemize}
\vspace{-0.9em}
\\\hline
\end{tabular}}
\vspace{0.05em}

\vspace{0.6em}{\color{blue}Call center access control and bank access control.}\vspace{0.6em}

\item Other comments?

\vspace{0.6em}{\color{blue}No.}\vspace{0.6em}

\end{itemize}

\subsection*{Basic Performance}
The following questions aim to offer an overall assessment of the service performance.

\vspace{1em} \noindent \textbf{Testing by service provider}

\begin{itemize}
\item Which datasets was the service tested on? (e.g., links to datasets that were used for testing, along with corresponding datasheets)

\vspace{0.6em}
\renewcommand{\arraystretch}{0.4}
{\small \begin{tabular}{|p{6.9cm}|}
\hline

\begin{itemize}
\item List the test datasets and provide links to these datasets.
\item Do the datasets have an associated datasheet? If yes, please attach.
\item Could these datasets be used for independent testing of the service? Did the data need to be changed or sampled before use? 
\end{itemize}
\vspace{-0.9em}
\\\hline
\end{tabular}}
\vspace{0.05em}

\vspace{0.6em}{\color{blue}FVC2002 DB1 (\url{http://bias.csr.unibo.it/fvc2002/databases.asp}).  This dataset does not have a datasheet. Yes, this dataset can be used for independent testing.}\vspace{0.6em}

\item Describe the testing methodology.

\vspace{0.6em}
\renewcommand{\arraystretch}{0.4}
{\small \begin{tabular}{|p{6.9cm}|}
\hline

\begin{itemize}
\item Please provide details on train, test and holdout data.
\item What performance metrics were used? (e.g. accuracy, error rates, AUC, precision/recall)
\item Please briefly justify the choice of metrics.
\end{itemize}
\vspace{-0.9em}
\\\hline
\end{tabular}}
\vspace{0.05em}

\vspace{0.6em}{\color{blue}Performance metrics were evaluated on a heldout set as specified by FVC2002. We used the same metrics as evaluated by FVC2002: equal error rate (EER), the lowest false non-match rate for a false match rate <= 1\% (FMR100), the lowest false non-match rate for a false match rate <= 0.1\% (FMR1000), the lowest false non-match rate for a false match rate = 0\% (ZeroFMR), number of rejected fingerprints during enrollment (RejENROLL), and number of rejected fingerprints during genuine and imposter matches (RejMATCH).}\vspace{0.6em}

\item Describe the test results.

\vspace{0.6em}
\renewcommand{\arraystretch}{0.4}
{\small \begin{tabular}{|p{6.9cm}|}
\hline

\begin{itemize}
\item Were latency, throughput, and availability measured?
\item If yes, briefly include those metrics as well.
\end{itemize}
\vspace{-0.9em}
\\\hline
\end{tabular}}
\vspace{0.05em}

\vspace{0.6em}{\color{blue}The accuracy results are as follows: EER = 3.7\%, FMR100 = 6.0\%, ZeroFMR = 12.4\%, RejENROLL = 0.0\%, RejMATCH = 0.0\%. We also measured average enrollment time: 0.14 sec and average matching time: 0.44 sec.}\vspace{0.6em}

\end{itemize}

\vspace{1em} \noindent \textbf{Testing by third parties}

\begin{itemize}
\item Is there a way to verify the performance metrics (e.g., via a service API )?

\vspace{0.6em}
\renewcommand{\arraystretch}{0.4}
{\small \begin{tabular}{|p{6.9cm}|}
\hline

\begin{itemize}
\item Briefly describe how a third party could independently verify the performance of the service.
\item Are there benchmarks publicly available and adequate for testing the service.
\end{itemize}
\vspace{-0.9em}
\\\hline
\end{tabular}}
\vspace{0.05em}

\vspace{0.6em}{\color{blue}Yes, a third party can call our service via the same RESTful API that our customers use.}\vspace{0.6em}

\item In addition to the service provider, was this service tested by any third party?

\vspace{0.6em}
\renewcommand{\arraystretch}{0.4}
{\small \begin{tabular}{|p{6.9cm}|}
\hline

\begin{itemize}
\item Please list all third parties that performed the testing.
\item Also, please include information about the tests and test results.
\end{itemize}
\vspace{-0.9em}
\\\hline
\end{tabular}}
\vspace{0.05em}

\vspace{0.6em}{\color{blue}No.}\vspace{0.6em}

\item Other comments?

\vspace{0.6em}{\color{blue}No.}\vspace{0.6em}

\end{itemize}

\subsection*{Safety}

The following questions aim to offer insights about potential unintentional harms, and mitigation efforts to eliminate or minimize those harms.

\vspace{1em} \noindent \textbf{General}

\begin{itemize}
\item Are you aware of possible examples of bias, ethical issues, or other safety risks as a result of using the service?

\vspace{0.6em}
\renewcommand{\arraystretch}{0.4}
{\small \begin{tabular}{|p{6.9cm}|}
\hline

\begin{itemize}
\item Were the possible sources of bias or unfairness analyzed?
\item Where do they arise from: the data? the particular techniques being implemented? other sources?
\item Is there any mechanism for redress if individuals are negatively affected?
\end{itemize}
\vspace{-0.9em}
\\\hline
\end{tabular}}
\vspace{0.05em}

\vspace{0.6em}{\color{blue}Yes, individuals with a history of manual labor will have poorer performance in fingerprint verification. Children will have poorer performance in fingerprint verification.  Individuals without fingerprints will be unable to use our service.  There is no mechanism for redress.}\vspace{0.6em}

\item Do you use data from or make inferences about individuals or groups of individuals. Have you obtained their consent?

\vspace{0.6em}
\renewcommand{\arraystretch}{0.4}
{\small \begin{tabular}{|p{6.9cm}|}
\hline

\begin{itemize}
\item How was it decided whose data to use or about whom to make inferences?
\item Do these individuals know that their data is being used or that inferences are being made about them? What were they told? When were they made aware? What kind of consent was needed from them? What were the procedures for gathering consent? Please attach the consent form to this declaration.
\item What are the potential risks to these individuals or groups? Might the service output interfere with individual rights? How are these risks being handled or minimized?
\item What trade-offs were made between the rights of these individuals and business interests?
\item Do they have the option to withdraw their data?  Can they opt out from inferences being made about them? What is the withdrawal procedure?
\end{itemize}
\vspace{-0.9em}
\\\hline
\end{tabular}}
\vspace{0.05em}

\vspace{0.6em}{\color{blue} Our training datasets come from international optical fingerprint databases available on the internet including FVC2000, FVC2002, and FVC2004. We did not do any further due diligence on these datasets.}\vspace{0.6em}

\end{itemize}

\noindent \textbf{Explainability}

\begin{itemize}
\item Are the service outputs explainable and/or interpretable?

\vspace{0.6em}
\renewcommand{\arraystretch}{0.4}
{\small \begin{tabular}{|p{6.9cm}|}
\hline

\begin{itemize}
\item Please explain how explainability is achieved (e.g. directly explainable algorithm, local explainability, explanations via examples).
\item Who is the target user of the explanation (ML expert, domain expert, general consumer, etc.)
\item Please describe any human validation of the explainability of the algorithms
\end{itemize}
\vspace{-0.9em}
\\\hline
\end{tabular}}
\vspace{0.05em}

\vspace{0.6em}{\color{blue}No.}\vspace{0.6em}

\end{itemize}

\vspace{1.2em} \noindent \textbf{Fairness}

\begin{itemize}
\item For each dataset used by the service: Was the dataset checked for bias? What efforts were made to ensure that it is fair and representative?

\vspace{0.6em}
\renewcommand{\arraystretch}{0.4}
{\small \begin{tabular}{|p{6.9cm}|}
\hline

\begin{itemize}
\item Please describe the data bias policies that were checked (such as with respect to known protected attributes), bias checking methods, and results (e.g., disparate error rates across different groups).
\item Was there any bias remediation performed on this dataset? Please provide details about the value of any bias estimates before and after it.
\item What techniques were used to perform the remediation? Please provide links to relevant technical papers.
\item How did the value of other performance metrics change as a result?
\end{itemize}
\vspace{-0.9em}
\\\hline
\end{tabular}}
\vspace{0.05em}

\vspace{0.6em}{\color{blue}We tested the service on a population of our company's employees and other office workers in our building, which includes younger and older adults, both male and female, with a range of skin tones.  We did not observe any systematic differential performance. No bias remediation was performed.}\vspace{0.6em}

\item Does the service implement and perform any bias detection and remediation?

\vspace{0.6em}
\renewcommand{\arraystretch}{0.4}
{\small \begin{tabular}{|p{6.9cm}|}
\hline

\begin{itemize} 
\item Please describe model bias policies that were checked, bias checking methods, and results (e.g., disparate error rates across different groups).
\item What procedures were used to perform the remediation? Please provide links or references to corresponding technical papers.
\item Please provide details about the value of any bias estimates before and after such remediation.
\item How did the value of other performance metrics change as a result?
\end{itemize}
\vspace{-0.9em}
\\\hline
\end{tabular}}
\vspace{0.05em}

\vspace{0.6em}{\color{blue}No.}\vspace{0.6em}

\end{itemize}

\noindent \textbf{Concept Drift}

\begin{itemize}

\item What is the expected performance on unseen data or data with different distributions?

\vspace{0.6em}
\renewcommand{\arraystretch}{0.4}
{\small \begin{tabular}{|p{6.9cm}|}
\hline

\begin{itemize}
\item Please describe any relevant testing done along with test results.
\end{itemize}

\vspace{-0.9em}
\\\hline
\end{tabular}}
\vspace{0.05em}

\vspace{0.6em}{\color{blue}Data from different types of sensors will result in extremely poor performance. Data from people from all working classes (those with frequent cuts on their hands) will result in a degradation of performance.}\vspace{0.6em}

\item Does your system make updates to its behavior based on newly ingested data?

\vspace{0.6em}
\renewcommand{\arraystretch}{0.4}
{\small \begin{tabular}{|p{6.9cm}|}
\hline

\begin{itemize}
\item Is the new data uploaded by your users? Is it generated by an automated process? Are the patterns in the data largely static or do they change over time?
\item Are there any performance guarantees/bounds?
\item Does the service have an automatic feedback/retraining loop, or is there a human in the loop?
\end{itemize}
\vspace{-0.9em}
\\\hline
\end{tabular}}
\vspace{0.05em}

\vspace{0.6em}{\color{blue}We have started to capture user data from our actual customer deployments and will retrain the algorithm including these additional images for v1.8.}\vspace{0.6em}

\item How is the service tested and monitored for model or performance drift over time?

\vspace{0.6em}
\renewcommand{\arraystretch}{0.4}
{\small \begin{tabular}{|p{6.9cm}|}
\hline

\begin{itemize}
\item If applicable, describe any relevant testing along with test results.
\end{itemize}

\vspace{-0.9em}
\\\hline
\end{tabular}}
\vspace{0.05em}

\vspace{0.6em}{\color{blue}Proprietary.}\vspace{0.6em}

\item How can the service be checked for correct, expected output when new data is added?

\vspace{0.6em}{\color{blue}We have not yet added any new data up to v1.7.}\vspace{0.6em}

\item Does the service allow for checking for differences between training and usage data?

\vspace{0.6em}
\renewcommand{\arraystretch}{0.4}
{\small \begin{tabular}{|p{6.9cm}|}
\hline

\begin{itemize}
\item Does it deploy mechanisms to alert the user of the difference?
\end {itemize}

\vspace{-0.9em}
\\\hline
\end{tabular}}
\vspace{0.05em}

\vspace{0.6em}{\color{blue}No.}\vspace{0.6em}

\item Do you test the service periodically?

\vspace{0.6em}
\renewcommand{\arraystretch}{0.4}
{\small \begin{tabular}{|p{6.9cm}|}
\hline

\begin{itemize}
\item Does the testing includes bias or fairness related aspects?
\item How has the value of the tested metrics evolved over time?
\end{itemize}

\vspace{-0.9em}
\\\hline
\end{tabular}}
\vspace{0.05em}

\vspace{0.6em}{\color{blue}Yes, we depute one of our staff members to visit our customer deployments once per quarter and do spot checks by enrolling and testing a few people. Metric evolution over time is confidential.}\vspace{0.6em}

\item Other comments?

\vspace{0.6em}{\color{blue}The general characteristics of fingerprints do not change over time.}\vspace{0.6em}

\end{itemize}

\subsection*{Security}

The following questions aim to assess the susceptibility to deliberate harms such as attacks by adversaries.

\begin{itemize}
\item How could this service be attacked or abused? Please describe.

\vspace{0.6em}{\color{blue}Many different attacks are possible.  Several are described in B. Biggio, G. Fumera, P. Russu,  L. Didaci, and F. Roli, ``Adversarial biometric recognition: A review on biometric system security from the adversarial machine-learning perspective,'' IEEE Signal Processing Magazine, vol. 32, no. 5, pp. 31--41, 2015.}\vspace{0.6em}

\item List applications or scenarios for which the service is not suitable.

\vspace{0.6em}
\renewcommand{\arraystretch}{0.4}
{\small \begin{tabular}{|p{6.9cm}|}
\hline

\begin{itemize}
\item Describe specific concerns and sensitive use cases.
\item Are there any procedures in place to ensure that the service will not be used for these applications?
\end{itemize}

\vspace{-0.9em}
\\\hline
\end{tabular}}
\vspace{0.05em}

\vspace{0.6em}{\color{blue}The service should not be used to investigate crimes, prosecute individuals, or used in any other way except for access control. There are no procedures in place to prevent such usage.}\vspace{0.6em}

\item How are you securing user or usage data?

\vspace{0.6em}
\renewcommand{\arraystretch}{0.4}
{\small \begin{tabular}{|p{6.9cm}|}
\hline

\begin{itemize}
\item Is usage data from service operations retained and stored?
\item How is the data being stored? For how long is the data stored?
\item Is user or usage data being shared outside the service? Who has access to the data?
\end{itemize}

\vspace{-0.9em}
\\\hline
\end{tabular}}
\vspace{0.05em}

\vspace{0.6em}{\color{blue}The usage data is not stored except when the user provides negative feedback and explicitly agrees for us to use current sample for retraining. The samples are deleted after retraining.}\vspace{0.6em}

\item Was the service checked for robustness against adversarial attacks?

\vspace{0.6em}
\renewcommand{\arraystretch}{0.4}
{\small \begin{tabular}{|p{6.9cm}|}
\hline

\begin{itemize}
\item Describe robustness policies that were checked, the type of attacks considered, checking methods, and results.
\end{itemize}

\vspace{-0.9em}
\\\hline
\end{tabular}}
\vspace{0.05em}

\vspace{0.6em}{\color{blue}We have checked for model-inversion and hill-climbing attacks using the techniques developed in M. Martinez-Diaz, J. Fierrez, J. Galbally, and J. Ortega-Garcia, ``An evaluation of indirect attacks and countermeasures in fingerprint verification systems,'' Pattern Recognition Letters, vol. 32, no. 12, pp. 1643--1651, 2011. Our service passed these tests.}\vspace{0.6em}

\item What is the plan to handle any potential security breaches?

\vspace{0.6em}
\renewcommand{\arraystretch}{0.4}
{\small \begin{tabular}{|p{6.9cm}|}
\hline

\begin{itemize}
\item Describe any protocol that is in place.
\end{itemize}

\vspace{-0.9em}
\\\hline
\end{tabular}}
\vspace{0.05em}

\vspace{0.6em}{\color{blue}We will shut down the service immediately in case of a potential security breach and only bring customers back online after site visits.}\vspace{0.6em}

\item Other comments?

\vspace{0.6em}{\color{blue}We take security very seriously.}\vspace{0.6em}

\end{itemize}

\subsection*{Lineage}

The following questions aim to overview how the service provider keeps track of details that might be required in the event of an audit by a third party, such as in the case of harm or suspicion of harm.

\vspace{1em} \noindent \textbf{Training Data}
\begin{itemize}
\item Does the service provide an as-is/canned model? Which datasets was the service trained on?

\vspace{0.6em}
\renewcommand{\arraystretch}{0.4}
{\small \begin{tabular}{|p{6.9cm}|}
\hline

\begin{itemize}
\item List the training datasets.
\item Where there any quality assurance processes employed while the data was collected or before use?
\item Were the datasets used for training built-for-purpose or were they re-purposed/adapted? Were the datasets created specifically for the purpose of training the models offered by this service?
\end{itemize}

\vspace{-0.9em}
\\\hline
\end{tabular}}
\vspace{0.05em}

\vspace{0.6em}{\color{blue}Our training datasets come from international optical fingerprint databases available on the internet including FVC2000, FVC2002, and FVC2004. All quality assurance was done by the dataset providers. They were purpose-built for the evaluation and testing of fingerprint verification systems.}\vspace{0.6em}

\item For each dataset: Are the training datasets publicly available?

\vspace{0.6em}
\renewcommand{\arraystretch}{0.4}
{\small \begin{tabular}{|p{6.9cm}|}
\hline

\begin{itemize}
\item Please provide a link to the datasets or the source of the datasets.
\end{itemize}

\vspace{-0.9em}
\\\hline
\end{tabular}}
\vspace{0.05em}

\vspace{0.6em}{\color{blue}Yes: \url{http://bias.csr.unibo.it/fvc2000/databases.asp}, \url{http://bias.csr.unibo.it/fvc2002/databases.asp}, \url{http://bias.csr.unibo.it/fvc2004/databases.asp}.}\vspace{0.6em}

\item For each dataset: Does the dataset have a datasheet or data statement?

\vspace{0.6em}
\renewcommand{\arraystretch}{0.4}
{\small \begin{tabular}{|p{6.9cm}|}
\hline

\begin{itemize} 
\item If available, attach the datasheet; otherwise, provide answers to questions from the datasheet as appropriate [to insert citation]
\end{itemize}

\vspace{-0.9em}
\\\hline
\end{tabular}}
\vspace{0.05em}

\vspace{0.6em}{\color{blue}No.}\vspace{0.6em}

\item Did the service require any transformation of the data in addition to those provided in the datasheet?

\vspace{0.6em}{\color{blue}No.}\vspace{0.6em}

\item Do you use synthetic data?

\vspace{0.6em}
\renewcommand{\arraystretch}{0.4}
{\small \begin{tabular}{|p{6.9cm}|}
\hline

\begin{itemize}
\item When? How was it created?
\item Briefly describe its properties and the creation procedure.

\end{itemize}

\vspace{-0.9em}
\\\hline
\end{tabular}}
\vspace{0.05em}

\vspace{0.6em}{\color{blue}No.}\vspace{0.6em}

\end{itemize}

\noindent \textbf{Trained Models}
\begin{itemize}
\item How were the models trained?

\vspace{0.6em}
\renewcommand{\arraystretch}{0.4}
{\small \begin{tabular}{|p{6.9cm}|}
\hline

\begin{itemize}
\item Please provide specific details (e.g., hyperparameters).
\end{itemize}

\vspace{-0.9em}
\\\hline
\end{tabular}}
\vspace{0.05em}

\vspace{0.6em}{\color{blue}Proprietary.}\vspace{0.6em}

\item When were the models last updated?

\vspace{0.6em}
\renewcommand{\arraystretch}{0.4}
{\small \begin{tabular}{|p{6.9cm}|}
\hline

\begin{itemize}
\item How much did the performance change with each update?
\item How often are the models retrained or updated?
\end{itemize}

\vspace{-0.9em}
\\\hline
\end{tabular}}
\vspace{0.05em}

\vspace{0.6em}{\color{blue}Proprietary.}\vspace{0.6em}

\item Did you use any prior knowledge or re-weight the data in any way before training?

\vspace{0.6em}{\color{blue}No.}\vspace{0.6em}

\item Other comments?

\vspace{0.6em}{\color{blue}No.}\vspace{0.6em}

\end{itemize}

\clearpage

\section{Sample FactSheet for a Trending Topics Service}
\label{trending}

\subsection*{Statement of purpose}

The following questions are aimed at providing an overview of the service provider and of the intended uses for the service. Valid answers include ``N/A'' (not applicable) and ``Proprietary'' (cannot be publicly disclosed, usually for competitive reasons).

\vspace{1.2em} \noindent \textbf{General}

\begin{itemize}
\item Who are ``you'' (the supplier) and what type of services do you typically offer (beyond this particular service)?

\vspace{0.6em}{\color{blue}DataTrendly specializes in natural language processing and time series analysis offering a wide range of products focused on the analysis of trending topics in several types of textual data, such as social media, news media, and scientific publications. }\vspace{0.6em}

\item What is this service about? 

\vspace{0.6em}
\renewcommand{\arraystretch}{0.4}
{\small \begin{tabular}{|p{6.9cm}|}
\hline 
\begin{itemize}
\item Briefly describe the service.
\item When was the service first released? When was the last release?
\item Who is the target user?
\end{itemize} 
\vspace{-0.9em}
\\\hline
\end{tabular}}
\vspace{0.05em}

\vspace{0.6em}{\color{blue}The DataTrendly's social media trending topics service allows our customers to check, identify, search for, and monitor trends on a variety of social media platforms. The service was first released in January 2014, and it was last updated in June 2018. Our target users are broad, anyone who wants to monitor a trending topic.  Some examples are a company wants to monitor its brand or media company that wants to model particular events.}\vspace{0.6em}

\item Describe the outputs of the service.

\vspace{0.6em}{\color{blue}The service is offered as a comprehensive set of RESTful API calls. The main API calls return a ranked list of N trending topics for a given time interval and a given list of key-phrases of interest.  For each trending topic it returns the key-phrases that triggered the topic, the time stamp of the topic, and a list of other related trending topics. }\vspace{0.6em}

\item What algorithms or techniques does this service implement?

\vspace{0.6em}
\renewcommand{\arraystretch}{0.4}
{\small \begin{tabular}{|p{6.9cm}|}
\hline
\begin{itemize}
\item Provide links to technical papers.
\end{itemize}
\vspace{-0.9em}
\\\hline
\end{tabular}}
\vspace{0.05em}

\vspace{0.6em}{\color{blue}    The service implements a mix of both commonly used techniques and our own proprietary techniques. Our users can specify what technique they want to use and can compare results from multiple techniques. 

\vspace{0.6em}For nowcasting and forecasting, the service implements known models like ARMA (Autoregressive, Moving Average), as well as proprietary neural network models. Our implementation of known techniques builds on Seabold and Perktold (2010) 

\vspace{0.6em}For past trends, it implements proprietary techniques for detecting sudden changes in time series data, which are tailored for social media data.

\vspace{0.6em}Seabold, Skipper, and Josef Perktold. ``Statsmodels: Econometric and statistical modeling with python.'' Proceedings of the 9th Python in Science Conference. 2010.

\vspace{0.6em}Link: \url{https://www.statsmodels.org/dev/tsa.htm}. 
}\vspace{0.6em}

\item What are the characteristics of the development team?

\vspace{0.6em}
\renewcommand{\arraystretch}{0.4}
{\small \begin{tabular}{|p{6.9cm}|}
\hline
\begin{itemize}
\item Do the teams charged with developing and maintaining this service reflect a diversity of opinions, backgrounds, and thought?
\end{itemize}
\vspace{-0.9em}
\\\hline
\end{tabular}}
\vspace{0.05em}

\vspace{0.6em}{\color{blue}Our team includes statisticians, AI researchers, developers, as well as a group of social scientists that help us evaluate the outputs of our service for the diverse use cases of our customers.  Our team is composed of individuals from a variety of socio-demographic backgrounds, with 32\% women and 11\% African American representation. }\vspace{0.6em}

\item Have you updated this FactSheet before?

\vspace{0.6em}
\renewcommand{\arraystretch}{0.4}
{\small \begin{tabular}{|p{6.9cm}|}
\hline
\begin{itemize}
\item When and how often?
\item What sections have changed?
\item Is the FactSheet updated every time the service is retrained or updated?
\end{itemize}
\vspace{-0.9em}
\\\hline
\end{tabular}}
\vspace{0.05em}

\vspace{0.6em}{\color{blue}We update the FactSheet with every service update, typically every 6 months. The following sections have changed in this FactSheet from the previous versions (available at URL): Statement of purpose (General), Basic performance, Safety (General, Fairness), and Lineage.}\vspace{0.6em}
\end{itemize}

\noindent \textbf{Usage}

\begin{itemize}
\item What is the intended use of the service output?

\vspace{0.6em}
\renewcommand{\arraystretch}{0.4}
{\small \begin{tabular}{|p{6.9cm}|}
\hline
\begin{itemize}
\item Briefly describe a simple use-case.
\end{itemize}
\vspace{-0.9em}
\\\hline
\end{tabular}}
\vspace{0.05em}

\vspace{0.6em}{\color{blue}Media organizations that wish to identify and monitor related topics to an event of interest form a common group of customers. Consider a sports magazine wanting to monitor trending topics related to the US Open tennis tournament. Monitoring only a few known key-phrases associated with the US Open, will likely miss topics that might be trending during the tournament. Our client can use our service to identify and monitor related topics that were found trending while the tournament was taking place. Our client can monitor trending topics within different time intervals and that exhibit different structural characteristics, such as topics that have gained sudden attention compared to those that have gained attention in a more incremental fashion. Our client can monitor and collate topics from multiple social media platforms. }\vspace{0.6em}

\item What are the key procedures followed while using the service?

\vspace{0.6em}
\renewcommand{\arraystretch}{0.4}
{\small \begin{tabular}{|p{6.9cm}|}
\hline
\begin{itemize}
\item How is the input provided? By whom?
\item How is the output returned?
\end{itemize}
\vspace{-0.9em}
\\\hline
\end{tabular}}
\vspace{0.05em}

\vspace{0.6em}{\color{blue}The client inputs a list of social media platforms that they wish to monitor for a specific set of key-phrases associated with the event or topic of interest to them. Using these, for each social media platform, our service returns a list of relevant trending topics. The client can then examine each of these topics to see what key-phrases are relevant and to identify other potentially related key-phrases, social media messages, and trending topics.}\vspace{0.6em}

\end{itemize}

\noindent \textbf{Domains and applications}
\begin{itemize}
\item What are the domains and applications the service was tested on or used for?

\vspace{0.6em}
\renewcommand{\arraystretch}{0.4}
{\small \begin{tabular}{|p{6.9cm}|}
\hline
\begin{itemize}
\item Were domain experts involved in the development, testing, and deployment? Please elaborate.
\end{itemize}
\vspace{-0.9em}
\\\hline
\end{tabular}}
\vspace{0.05em}

\vspace{0.6em}{\color{blue}Our service is often being used for brand monitoring on social media and for event monitoring by various local and regional media companies, where we work very closely with our customers as they understand better the subtleties around their brands and/or they often have a better grasp of the context of the events they are interested in monitoring.}\vspace{0.6em}

\item How is the service being used by your customers or users?

\vspace{0.6em}
\renewcommand{\arraystretch}{0.4}
{\small \begin{tabular}{|p{6.9cm}|}
\hline
\begin{itemize}
\item Are you enabling others to build a solution by providing a cloud service or is your application end-user facing?
\item Is the service output used as-is, is it fed directly into another tool or actuator, or is there human input/oversight before use?
\item Do users rely on pre-trained/canned models or can they train their own models?
\item Do your customers typically use your service in a time critical setup (e.g. they have limited time to evaluate the output)? Or do they incorporate it in a slower decision making process? Please elaborate.
\end{itemize}
\vspace{-0.9em}
\\\hline
\end{tabular}}
\vspace{0.05em}

\vspace{0.6em}{\color{blue}Our service is typically integrated by our customers in their own in-house data monitoring, data gathering platforms, but can also be integrated into more application-tailored solutions by our customers. We encourage our customers to qualitatively validate the output of the service, and we work closely with them as they integrate our service and use it. For our media customers, the outputs of our service help them better identify and contextualize the stories they cover, while our brand customers use them for marketing decisions and crisis management purposes.}\vspace{0.6em}

\item List applications that the service has been used for in the past.

\vspace{0.6em}
\renewcommand{\arraystretch}{0.4}
{\small \begin{tabular}{|p{6.9cm}|}
\hline
\begin{itemize}
\item Please provide information about these applications or relevant pointers.
\item Please provide key performance results for those applications.
\end{itemize}
\vspace{-0.9em}
\\\hline
\end{tabular}}
\vspace{0.05em}

\vspace{0.6em}{\color{blue}Our service is used for brand monitoring and event monitoring. The specific details of each application where the service was used is customer-proprietary information.}\vspace{0.6em}

\item Other comments?

\vspace{0.6em}{\color{blue}N/A.}\vspace{0.6em}

\end{itemize}

\subsection*{Basic Performance}
The following questions aim to offer an overall assessment of the service performance.

\vspace{1em} \noindent \textbf{Testing by service provider}

\begin{itemize}
\item Which datasets was the service tested on? (e.g., links to datasets that were used for testing, along with corresponding datasheets)

\vspace{0.6em}
\renewcommand{\arraystretch}{0.4}
{\small \begin{tabular}{|p{6.9cm}|}
\hline

\begin{itemize}
\item List the test datasets and provide links to these datasets.
\item Do the datasets have an associated datasheet? If yes, please attach.
\item Could these datasets be used for independent testing of the service? Did the data need to be changed or sampled before use? 
\end{itemize}
\vspace{-0.9em}
\\\hline
\end{tabular}}
\vspace{0.05em}

\vspace{0.6em}{\color{blue}    We release test results and benchmarks at the end of the year based on a set of popular trending topics that year and our predictions of how those trends will fare at different time stamps.  Data about these trends and comprehensive result reports are available at \url{https://datatrendly.com/reports/}.

\vspace{0.6em}Yes, we provide a datasheet for each end-of-year popular trending topics release. Each release contains the set of popular key-phrases for that year, timeseries corresponding to their popularity, and our prediction at different timestamps. See them attached at the end of this FactSheet.

\vspace{0.6em}These datasets can be used to further analyze our performance, but also to check it on our platform which allows retrospective browsing. In addition, our customers can check our predictions retrospectively for any other trend of interest.}\vspace{0.6em}

\item Describe the testing methodology.

\vspace{0.6em}
\renewcommand{\arraystretch}{0.4}
{\small \begin{tabular}{|p{6.9cm}|}
\hline

\begin{itemize}
\item Please provide details on train, test and holdout data.
\item What performance metrics were used? (e.g. accuracy, error rates, AUC, precision/recall)
\item Please briefly justify the choice of metrics.
\end{itemize}
\vspace{-0.9em}
\\\hline
\end{tabular}}
\vspace{0.05em}

\vspace{0.6em}{\color{blue}The performance metrics include: (1) scale dependent error rates that are often useful per case study basis, such as mean absolute error (MAE) and root mean square errors (RMSE);  (2) percentage error rates as they allow for a better comparison of results between different timeseries, such as mean absolute percentage error (MAPE) and its symmetric version Armstrong (1978, p. 348), and (3) and scaled error rates that are preferred for comparisons of timeseries originating from different platforms or of different nature, explained in Hyndman \& Koehler (2006). In addition, we also examine the residuals for any systematic trends, and that there is no correlation between residuals for which we use the Box-Pierce and Ljung-Box tests. 

\vspace{0.6em}Hyndman, R. J., \& Koehler, A. B. (2006). Another look at measures of forecast accuracy. International Journal of Forecasting, 22, 679--688. \url{https://robjhyndman.com/publications/automatic-forecasting/}

\vspace{0.6em}Armstrong, J. S. (1978). Long-range forecasting: From crystal ball to computer. John Wiley \& Sons.}\vspace{0.6em}

\item Describe the test results.

\vspace{0.6em}
\renewcommand{\arraystretch}{0.4}
{\small \begin{tabular}{|p{6.9cm}|}
\hline

\begin{itemize}
\item Were latency, throughput, and availability measured?
\item If yes, briefly include those metrics as well.
\end{itemize}
\vspace{-0.9em}
\\\hline
\end{tabular}}
\vspace{0.05em}

\vspace{0.6em}{\color{blue}Our service makes real time predictions on how various trends may fare, and it is critical for us that our customers get access to the information they need timely and reliably.  As a result, latency, throughput, and availability are critical metrics for a service like ours. For instance, our maximum latency to answer a query (between receiving a query and producing a result) is 2s.}\vspace{0.6em}

\end{itemize}

\vspace{1em} \noindent \textbf{Testing by third parties}

\begin{itemize}
\item Is there a way to verify the performance metrics (e.g., via a service API )?

\vspace{0.6em}
\renewcommand{\arraystretch}{0.4}
{\small \begin{tabular}{|p{6.9cm}|}
\hline

\begin{itemize}
\item Briefly describe how a third party could independently verify the performance of the service.
\item Are there benchmarks publicly available and adequate for testing the service.
\end{itemize}
\vspace{-0.9em}
\\\hline
\end{tabular}}
\vspace{0.05em}

\vspace{0.6em}{\color{blue}Yes, to some extent the performance metrics can be independently verified by third parties, if those third parties are our customers. Otherwise, it can be done only based on the data we release at the end of the year on popular trends that year and our corresponding predictions at different timestamps. For confidentiality and business reasons we do not allow third parties access to the work we do for our customers. }\vspace{0.6em}

\item In addition to the service provider, was this service tested by any third party?

\vspace{0.6em}
\renewcommand{\arraystretch}{0.4}
{\small \begin{tabular}{|p{6.9cm}|}
\hline

\begin{itemize}
\item Please list all third parties that performed the testing.
\item Also, please include information about the tests and test results.
\end{itemize}
\vspace{-0.9em}
\\\hline
\end{tabular}}
\vspace{0.05em}

\vspace{0.6em}{\color{blue}No.}\vspace{0.6em}

\item Other comments?

\vspace{0.6em}{\color{blue}No.}\vspace{0.6em}

\end{itemize}

\subsection*{Safety}

The following questions aim to offer insights about potential unintentional harms, and mitigation efforts to eliminate or minimize those harms.

\vspace{1em} \noindent \textbf{General}

\begin{itemize}
\item Are you aware of possible examples of bias, ethical issues, or other safety risks as a result of using the service?

\vspace{0.6em}
\renewcommand{\arraystretch}{0.4}
{\small \begin{tabular}{|p{6.9cm}|}
\hline

\begin{itemize}
\item Were the possible sources of bias or unfairness analyzed?
\item Where do they arise from: the data? the particular techniques being implemented? other sources?
\item Is there any mechanism for redress if individuals are negatively affected?
\end{itemize}
\vspace{-0.9em}
\\\hline
\end{tabular}}
\vspace{0.05em}

\vspace{0.6em}{\color{blue}    We are not aware of broad ethical issues concerning our service. Some ethical issues may arise in the context of more sensitive applications such as identifying and monitoring trends related to anti-governmental movements. Although our service is not centered around identifying or making inference about individuals, one could use the trending topics to identify those posting about them. In this particular case, while we have a policy to not engage in such use-cases (policy that our customers are made aware of), our customers can use our service to monitor key-phrases beyond the close engagements we have with them. When we have any suspicion about the topics being monitored  via our service, we block the usage of the related key-phrases. 

\vspace{0.6em}In addition to such concerns, various biases in the data might skew the interpretation of the output we provide. Social media data is known not to be representative, and different social media platforms might exhibit different representational biases. The characteristics of each platform might also influence how users are likely to behave, such as what content they are likely to share. These biases may also evolve over time depending on seasonal patterns, external circumstances, and because of changes in the user-base or in the features of each social media platform.  These affect the type of insights that our customers can draw from the data and the trending topics we identify. For a comprehensive overview of data biases that our data tends to suffer from, see Olteanu et al. (2016, p. 6). We discuss these issues in detail with our customers, including how they might impact their analysis. We recommend qualitative analyses of the outputs, as well as tracking the same trends across multiple platforms. 

\vspace{0.6em}Olteanu et al. ``Social data: Biases, methodological pitfalls, and ethical boundaries.'' SSRN (2016).}\vspace{0.6em}

\item Do you use data from or make inferences about individuals or groups of individuals. Have you obtained their consent?

\vspace{0.6em}
\renewcommand{\arraystretch}{0.4}
{\small \begin{tabular}{|p{6.9cm}|}
\hline

\begin{itemize}
\item How was it decided whose data to use or about whom to make inferences?
\item Do these individuals know that their data is being used or that inferences are being made about them? What were they told? When were they made aware? What kind of consent was needed from them? What were the procedures for gathering consent? Please attach the consent form to this declaration.
\item What are the potential risks to these individuals or groups? Might the service output interfere with individual rights? How are these risks being handled or minimized?
\item What trade-offs were made between the rights of these individuals and business interests?
\item Do they have the option to withdraw their data?  Can they opt out from inferences being made about them? What is the withdrawal procedure?
\end{itemize}
\vspace{-0.9em}
\\\hline
\end{tabular}}
\vspace{0.05em}

\vspace{0.6em}{\color{blue}No, our service is not centered around making inferences about individuals or groups of individuals. Any analysis we make is content based, not user based. Given that we use only public data, we do not obtain explicit consent from the users of these platforms. However, some topics might be of interest to certain groups, and we acknowledge that in certain use-cases (as mentioned above) this may lead to safety concerns. To minimize such risks, whenever there is a suspicion of such a use-case we block the use of related key-phrases on our service.}\vspace{0.6em}

\end{itemize}

\noindent \textbf{Explainability}

\begin{itemize}
\item Are the service outputs explainable and/or interpretable?

\vspace{0.6em}
\renewcommand{\arraystretch}{0.4}
{\small \begin{tabular}{|p{6.9cm}|}
\hline

\begin{itemize}
\item Please explain how explainability is achieved (e.g. directly explainable algorithm, local explainability, explanations via examples).
\item Who is the target user of the explanation (ML expert, domain expert, general consumer, etc.)
\item Please describe any human validation of the explainability of the algorithms
\end{itemize}
\vspace{-0.9em}
\\\hline
\end{tabular}}
\vspace{0.05em}

\vspace{0.6em}{\color{blue}We do not provide explicit explanations for our inferences. }\vspace{0.6em}

\end{itemize}

\vspace{1.2em} \noindent \textbf{Fairness}

\begin{itemize}
\item For each dataset used by the service: Was the dataset checked for bias? What efforts were made to ensure that it is fair and representative?

\vspace{0.6em}
\renewcommand{\arraystretch}{0.4}
{\small \begin{tabular}{|p{6.9cm}|}
\hline

\begin{itemize}
\item Please describe the data bias policies that were checked (such as with respect to known protected attributes), bias checking methods, and results (e.g., disparate error rates across different groups).
\item Was there any bias remediation performed on this dataset? Please provide details about the value of any bias estimates before and after it.
\item What techniques were used to perform the remediation? Please provide links to relevant technical papers.
\item How did the value of other performance metrics change as a result?
\end{itemize}
\vspace{-0.9em}
\\\hline
\end{tabular}}
\vspace{0.05em}

\vspace{0.6em}{\color{blue}Although we report known statistics about the socio-demographic composition of each social media platform we work with our customers and discuss with them about the type of conclusions they can draw from the outputs of our system, we do not perform our own bias checks.}\vspace{0.6em}

\item Does the service implement and perform any bias detection and remediation?

\vspace{0.6em}
\renewcommand{\arraystretch}{0.4}
{\small \begin{tabular}{|p{6.9cm}|}
\hline

\begin{itemize} 
\item Please describe model bias policies that were checked, bias checking methods, and results (e.g., disparate error rates across different groups).
\item What procedures were used to perform the remediation? Please provide links or references to corresponding technical papers.
\item Please provide details about the value of any bias estimates before and after such remediation.
\item How did the value of other performance metrics change as a result?
\end{itemize}
\vspace{-0.9em}
\\\hline
\end{tabular}}
\vspace{0.05em}

\vspace{0.6em}{\color{blue}No.}\vspace{0.6em}

\end{itemize}

\noindent \textbf{Concept Drift}

\begin{itemize}

\item What is the expected performance on unseen data or data with different distributions?

\vspace{0.6em}
\renewcommand{\arraystretch}{0.4}
{\small \begin{tabular}{|p{6.9cm}|}
\hline

\begin{itemize}
\item Please describe any relevant testing done along with test results.
\end{itemize}

\vspace{-0.9em}
\\\hline
\end{tabular}}
\vspace{0.05em}

\vspace{0.6em}{\color{blue}N/A. We build a different model for each query submitted by our customers in real-time.}\vspace{0.6em}

\item Does your system make updates to its behavior based on newly ingested data?

\vspace{0.6em}
\renewcommand{\arraystretch}{0.4}
{\small \begin{tabular}{|p{6.9cm}|}
\hline

\begin{itemize}
\item Is the new data uploaded by your users? Is it generated by an automated process? Are the patterns in the data largely static or do they change over time?
\item Are there any performance guarantees/bounds?
\item Does the service have an automatic feedback/retraining loop, or is there a human in the loop?
\end{itemize}
\vspace{-0.9em}
\\\hline
\end{tabular}}
\vspace{0.05em}

\vspace{0.6em}{\color{blue}N/A.}\vspace{0.6em}

\item How is the service tested and monitored for model or performance drift over time?

\vspace{0.6em}
\renewcommand{\arraystretch}{0.4}
{\small \begin{tabular}{|p{6.9cm}|}
\hline

\begin{itemize}
\item If applicable, describe any relevant testing along with test results.
\end{itemize}

\vspace{-0.9em}
\\\hline
\end{tabular}}
\vspace{0.05em}

\vspace{0.6em}{\color{blue}Not purposefully. However, we keep track of our performance metrics for each use-case and application our service is used for over time. This also allows us to check for potential variations in performance over time as the characteristics of the data might vary with factors specific to each social media platform.}\vspace{0.6em}

\item How can the service be checked for correct, expected output when new data is added?

\vspace{0.6em}{\color{blue}N/A.}\vspace{0.6em}

\item Does the service allow for checking for differences between training and usage data?

\vspace{0.6em}
\renewcommand{\arraystretch}{0.4}
{\small \begin{tabular}{|p{6.9cm}|}
\hline

\begin{itemize}
\item Does it deploy mechanisms to alert the user of the difference?
\end {itemize}

\vspace{-0.9em}
\\\hline
\end{tabular}}
\vspace{0.05em}

\vspace{0.6em}{\color{blue}N/A.}\vspace{0.6em}

\item Do you test the service periodically?

\vspace{0.6em}
\renewcommand{\arraystretch}{0.4}
{\small \begin{tabular}{|p{6.9cm}|}
\hline

\begin{itemize}
\item Does the testing includes bias or fairness related aspects?
\item How has the value of the tested metrics evolved over time?
\end{itemize}

\vspace{-0.9em}
\\\hline
\end{tabular}}
\vspace{0.05em}

\vspace{0.6em}{\color{blue}Yes, as mentioned above, we keep track of our performance metrics for each use-case and application our service is used for. }\vspace{0.6em}

\item Other comments?

\vspace{0.6em}{\color{blue}No.}\vspace{0.6em}

\end{itemize}

\subsection*{Security}

The following questions aim to assess the susceptibility to deliberate harms such as attacks by adversaries.

\begin{itemize}
\item How could this service be attacked or abused? Please describe.

\vspace{0.6em}{\color{blue}Our service is used in a subscription mode, where the user registers and subscribes to a custom set of trending topics. This leaves us in control of the traffic and hence secure from attacks like those on typical SQL query servers.}\vspace{0.6em}

\item List applications or scenarios for which the service is not suitable.

\vspace{0.6em}
\renewcommand{\arraystretch}{0.4}
{\small \begin{tabular}{|p{6.9cm}|}
\hline

\begin{itemize}
\item Describe specific concerns and sensitive use cases.
\item Are there any procedures in place to ensure that the service will not be used for these applications?
\end{itemize}

\vspace{-0.9em}
\\\hline
\end{tabular}}
\vspace{0.05em}

\vspace{0.6em}{\color{blue}We have procedures in place to disallow subscriptions by customers interested in monitoring ethically questionable topics such as hate speech or pornographic content.}\vspace{0.6em}

\item How are you securing user or usage data?

\vspace{0.6em}
\renewcommand{\arraystretch}{0.4}
{\small \begin{tabular}{|p{6.9cm}|}
\hline

\begin{itemize}
\item Is usage data from service operations retained and stored?
\item How is the data being stored? For how long is the data stored?
\item Is user or usage data being shared outside the service? Who has access to the data?
\end{itemize}

\vspace{-0.9em}
\\\hline
\end{tabular}}
\vspace{0.05em}

\vspace{0.6em}{\color{blue}Yes, usage data is stored per tenant for a period of 2 weeks. The data is used for understanding usage patterns, helpful for improving the service. Only our own data science team has access to this data. The data is stored in encrypted format on our servers.}\vspace{0.6em}

\item Was the service checked for robustness against adversarial attacks?

\vspace{0.6em}
\renewcommand{\arraystretch}{0.4}
{\small \begin{tabular}{|p{6.9cm}|}
\hline

\begin{itemize}
\item Describe robustness policies that were checked, the type of attacks considered, checking methods, and results.
\end{itemize}

\vspace{-0.9em}
\\\hline
\end{tabular}}
\vspace{0.05em}

\vspace{0.6em}{\color{blue}Yes. We do have algorithms in place to discard ``wrong'' feedback data which may deteriorate the performance of the service. }\vspace{0.6em}

\item What is the plan to handle any potential security breaches?

\vspace{0.6em}
\renewcommand{\arraystretch}{0.4}
{\small \begin{tabular}{|p{6.9cm}|}
\hline

\begin{itemize}
\item Describe any protocol that is in place.
\end{itemize}

\vspace{-0.9em}
\\\hline
\end{tabular}}
\vspace{0.05em}

\vspace{0.6em}{\color{blue}Give the short retention period of customer data and the nature of our service, we believe that there is both a low risk of security breaches and they have limited ramifications. If those happen, and limited customer data is compromised or leaked, we will notify anyone affected immediately.}\vspace{0.6em}

\item Other comments?

\vspace{0.6em}{\color{blue}No.}\vspace{0.6em}

\end{itemize}

\subsection*{Lineage}

The following questions aim to overview how the service provider keeps track of details that might be required in the event of an audit by a third party, such as in the case of harm or suspicion of harm.

\vspace{1em} \noindent \textbf{Training Data}
\begin{itemize}
\item Does the service provide an as-is/canned model? Which datasets was the service trained on?

\vspace{0.6em}
\renewcommand{\arraystretch}{0.4}
{\small \begin{tabular}{|p{6.9cm}|}
\hline

\begin{itemize}
\item List the training datasets.
\item Where there any quality assurance processes employed while the data was collected or before use?
\item Were the datasets used for training built-for-purpose or were they re-purposed/adapted? Were the datasets created specifically for the purpose of training the models offered by this service?
\end{itemize}

\vspace{-0.9em}
\\\hline
\end{tabular}}
\vspace{0.05em}

\vspace{0.6em}{\color{blue}N/A. We build a different model for each use-case and trend. We generate our time series based on the queries submitted by our customers; thus, they are built and used for the purpose for which they were generated. }\vspace{0.6em}

\item For each dataset: Are the training datasets publicly available?

\vspace{0.6em}
\renewcommand{\arraystretch}{0.4}
{\small \begin{tabular}{|p{6.9cm}|}
\hline

\begin{itemize}
\item Please provide a link to the datasets or the source of the datasets.
\end{itemize}

\vspace{-0.9em}
\\\hline
\end{tabular}}
\vspace{0.05em}

\vspace{0.6em}{\color{blue}No. We only release data and results corresponding to the most popular trends at the end of the year. We do not release data corresponding to customers' engagements. We do prepare datasheets for every data release.}\vspace{0.6em}

\item For each dataset: Does the dataset have a datasheet or data statement?

\vspace{0.6em}
\renewcommand{\arraystretch}{0.4}
{\small \begin{tabular}{|p{6.9cm}|}
\hline

\begin{itemize} 
\item If available, attach the datasheet; otherwise, provide answers to questions from the datasheet as appropriate [to insert citation]
\end{itemize}

\vspace{-0.9em}
\\\hline
\end{tabular}}
\vspace{0.05em}

\vspace{0.6em}{\color{blue}N/A.}\vspace{0.6em}

\item Did the service require any transformation of the data in addition to those provided in the datasheet?

\vspace{0.6em}{\color{blue}N/A.}\vspace{0.6em}

\item Do you use synthetic data?

\vspace{0.6em}
\renewcommand{\arraystretch}{0.4}
{\small \begin{tabular}{|p{6.9cm}|}
\hline

\begin{itemize}
\item When? How was it created?
\item Briefly describe its properties and the creation procedure.

\end{itemize}

\vspace{-0.9em}
\\\hline
\end{tabular}}
\vspace{0.05em}

\vspace{0.6em}{\color{blue}No.}\vspace{0.6em}

\end{itemize}

\noindent \textbf{Trained Models}
\begin{itemize}
\item How were the models trained?

\vspace{0.6em}
\renewcommand{\arraystretch}{0.4}
{\small \begin{tabular}{|p{6.9cm}|}
\hline

\begin{itemize}
\item Please provide specific details (e.g., hyperparameters).
\end{itemize}

\vspace{-0.9em}
\\\hline
\end{tabular}}
\vspace{0.05em}

\vspace{0.6em}{\color{blue}    To make a forecast for a given time slot in the future and a given trend of interest, we use the historical behavior of the time series corresponding to this trend, as well as other historical information from related trends on a given social media platform or from several platforms, which can be either automatically extracted or hand selected by experts in our customer teams, or a combination of both.

\vspace{0.6em}We build a different model for each use-case and trend.}\vspace{0.6em}

\item When were the models last updated?

\vspace{0.6em}
\renewcommand{\arraystretch}{0.4}
{\small \begin{tabular}{|p{6.9cm}|}
\hline

\begin{itemize}
\item How much did the performance change with each update?
\item How often are the models retrained or updated?
\end{itemize}

\vspace{-0.9em}
\\\hline
\end{tabular}}
\vspace{0.05em}

\vspace{0.6em}{\color{blue}N/A.}\vspace{0.6em}

\item Did you use any prior knowledge or re-weight the data in any way before training?

\vspace{0.6em}{\color{blue}In some cases, yes, we do. We incorporate in our models historical information about time series selected by domain experts from our client teams that are expected to share some relationship with the trends of interests.}\vspace{0.6em}

\item Other comments?

\vspace{0.6em}{\color{blue}No.}\vspace{0.6em}

\end{itemize}

\end{document}